# Information Measure as Time Complexity

GUO ZHAO, Southwest University for Nationalities

Abstract: The concept of Shannon entropy of random variables was generalized to measurable functions in general, and to simple functions with finite values in particular. It is shown that the information measure of a function is related to the time complexity of search problems concerning the function in question. Formally, given a **Turing reduction** from a search problem $f(x) = y$ to another function $g(x)$, the average amount of information about $f(x) = y$ provided by each query is equal to the average mutual information $\mathrm{I}(f; g)$. As a result, the **average** number of queries is $\dfrac{\mathrm{I}(f = y)}{\mathrm{I}(f; g)}$, where $\mathrm{I}(f = y)$ is the amount of self-information about the event $\{f = y\}$. In the idea case, if $\dfrac{\mathrm{I}(f = y)}{\mathrm{I}(f; g)}$ is polynomial in the size of input and the function $g(x)$ can be computed in polynomial time, then the problem $f(x) = y$ also has polynomial-time algorithm. As it turns out, our information-based complexity estimation is a natural setting in which to study the power of randomized or probabilistic algorithms. Applying to decision problems, our result provides a strong evidence that $P = RP = BPP$. Using brute-force search as a benchmark for efficiency, our results also support that $P \neq PP$. Further, applying an argument similar to Carnot's work on measuring the efficiency of an ideal heat engine in thermodynamics, it is shown that $\dfrac{\mathrm{I}(f = y)}{\mathrm{H}(f)}$ is a **lower bound** on query complexity of solving $f(x) = y$. According to Markov's inequality, if $\dfrac{\mathrm{I}(f = y)}{\mathrm{H}(f)}$ is exponential in the size of input, then the probability of $f(x) = y$ having polynomial-time algorithm is a negative exponential in the size of input. This result enables us to reduce the problem of proving $FP \neq FNP$ to the computation of the entropy of certain simple functions, such as the subset sum function. As a result, $P \neq NP$ is provable, answering a long-standing open problem.

---

This work is supported by ...

Author's addresses: G. Zhao, College of Computer Science and Technology, Southwest University for Nationalities, Chengdu 610041, China. E-mail: zhaoguo@swun.cn

## 1. INTRODUCTION

This paper addresses the following questions: How much information is contained in a function? Or more specifically, given a measurable function $f : X \to \Re$ and a random input $x \in X$, how much information is needed to specify the exact output $f(x)$? Further, given a *Turing reduction* from given search problem $f(x) = y$ to another search problem $g(x) = z$ (see Goldreich [2008]; Ding-Zhu Du, Ker-I Ko [2013]), how much information about $f(x) = y$ can be obtained from querying the solution(s) of $g(x) = z$? And what is the **least** number of queries required to solve $f(x) = y$? [1]

In fact, the same idea has been used by Shannon [1949] to analyze the practical secrecy of cryptosystem. Using functional notation, denote the enciphering function to be $y = f_K(x)$, where $K$ corresponds to the particular key being used. Given an enciphered message ($y$), any system can be broken, at least in theory, by merely trying each possible key ($K$) until the most possible solution ($x$) is obtained. The complexity of the system can be measured by the **least** number of trials to carry out the solution. Suppose that the entropy of key space is given by $\mathrm{H}(K)$, which is the average amount of information associated with the choice of keys. If each trial has $S$ *equally possible* results then the **least** number of trials will be $\dfrac{\mathrm{H}(K)}{\log S}$. However, if the key space has very different probabilities, then only a small amount of information is obtained from testing a single key in complete trial and error.

Further, Shannon [1949] pointed out that the time complexity of decoding a cryptosystem is similar to the coin weighing problem (see also Erdos and Renyi [1963]). A typical example is the following: if one coin in $27$ is slightly lighter than the rest, what is the **least** number of weighings required to isolate it using a chemist's balance? The correct answer is $3$, obtained by first dividing the coin into three groups of $9$ each. The set of coins corresponds to the set of keys, the counterfeit coin to the correct key, and the weighing procedure to a trial or test. The original uncertainty is $\log 27$ bits, and each trial yields $\log 3$ bits of information; thus, at least $\dfrac{\log 27}{\log 3} = 3$ trials are required.

There is an analogy of this problem in physics. To raise an object of mass $m$ upward a vertical height $h$, the work done against the gravitational force can *not* be less than $mgh$, where $g$ is the gravitational acceleration (see Feynman *et al.* [2013]). The situations of our questions are essentially the same. But the question we now face is how to measure the amount of information contained in given function (for similar ideas see Packel and Traub [1987]; Yao, A.C. [1988]; Shah and Sharma [2000] ; Tadao Takaoka, Yuji Nakagawa [2010]).

---

[1] If $f(x)$ happens to be **continuous**, then only *partial* information is available and we can solve the search problem $f(x) = y$ only approximately. In this case, the degree of accuracy must be considered. As a result, the computational complexity of solving the search problem $f(x) = y$ will depend on the error rate. Research in this spirit is called Information-based Complexity. For details see Packel and Traub [1987]. As for the efficiency of algorithms from "continuous" classical mathematics, see Smale [1985]. Historically, the idea of the development of computational complexity theory for continuous mathematics at least dated back to Von Neumann, who said: ".. .a detailed, highly mathematical and more specifically analytical, theory of automata and of information is needed"( Von Neumann [1963]).

To this end, the concept of Shannon entropy of random variables was extended to measurable functions in general, and to simple functions with finite values in particular (see Patrick Billingsley [1995]). Formally, let $(X, \mu)$ be a measurable space with measure $\mu(X) < \infty$ and $f : X \to \Re$ be a measurable function. The key point is to approximate $f : X \to \Re$ by a simple function with finite values, which is obtained by partitioning the range of $f(x)$ in the same way as computing its Lebesgue integral (see Patrick Billingsley [1995]). As a result, these values, coupled with the probabilities of occurrence of their preimages, form a random variable in itself (see Fig. 1 and section 2 for details). So, it is natural to approximate the entropy of the given function by the entropy of the resulting random variable.

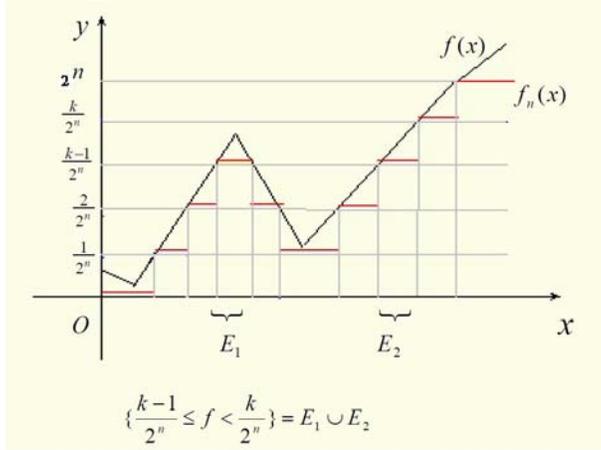

Fig. 1. Approximating a measurable function by simple functions.

Specially, if $f : X \to \Re$ is a simple function with values $y_1, y_2, \cdots, y_n$, then the self-information associated with the event $\{f = y_k\}$ is equal to $\mathrm{I}(f = y_k) = -\log p_k$, where $p_k = \mathrm{Pr}(f = y_k) = \dfrac{\mu(f^{-1}(y_k))}{\mu(X)}$. Consequently, the entropy of $f$ is defined to be $\mathrm{H}(f) = -\sum_{k=1}^{n} p_k \log p_k$. If the measure $\mu(X) = 1$, then $f : X \to \Re$ itself becomes into a random variable, and our definition of entropy coincides with that of Shannon [1948].

However, for cryptographic purposes, especially in public-key cryptography, the given cipher text $y_k = f(x)$ may convey some information about the corresponding preimage(s). As such, it may be possible that one can decrypt *one* cipher text $y_k$ without inverting the enciphering function $f : X \to \Re$ directly. Intuitively, decrypting *one* cipher text $y_k$ may not be computationally equivalent to inverting the enciphering function $f : X \to \Re$, which means decrypting *all* cipher texts (see Diffie and Hellman [1976]).

To overcome this logical difficulty, we have to make good use of the information conveyed by the given search problem $f(x) = y_k$. In some idea cases, the information hidden in the function $f : X \to \Re$ and the information about the value $y_k$ may enable us to reduce the search problem $f(x) = y_k$ to another search problem $g(x) = z_i$. Throughout, we assume that the function $g : Y \to \Re$ is always measurable and hence can be approximated by a simple function with finite values $z_0, z_1, \cdots, z_m$.

A typical example is the optimization problem of real-valued functions. Given a *differentiable* function $y = f(x)$ on a closed interval $[a,b]$, Fermat's theorem states that if $f(x)$ has a local extremum at $x_0 \in (a,b)$, then its derivative $f'(x)$ must satisfy $f'(x_0) = 0$. By using Fermat's theorem, the local extremum of $f(x)$ is found by solving an equation $f'(x_0) = 0$. As a result, the information of differentiability of $f(x)$ and optimality of $f(x_0)$ enables us to reduce the optimization problem $\min_{x \in (a,b)} f(x)$ to another search problem $f'(x) = 0$ (see Fig. 2 and section 3.4 for details).

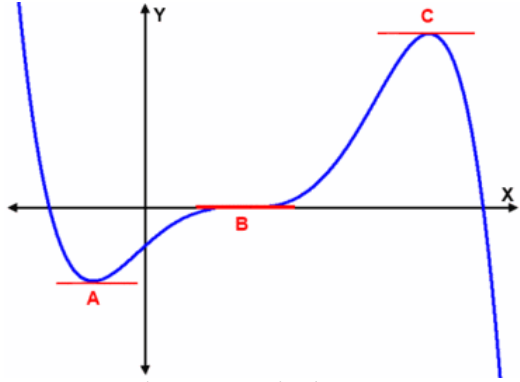

Fig. 2. Fermat's Theorem.

Formally, assume that there is a *Turing reduction* from given search problem $f(x) = y_k$ to another search problem $g(x) = z_i$, denoted by $f = y_k \leq_T g = z_i$ (see Ding-Zhu Du, Ker-I Ko [2013]). [2] This means that there exists an algorithm that solves $f(x) = y_k$ by calling as a *subroutine* another algorithm that solves $g(x) = z_i$ (see Goldreich [2008]). Further, if each query can be done in

---

polynomial time, then to estimate the time complexity of solving $f(x) = y_k$ we just need to estimate the **least** number of queries, denoted as $time(f = y_k)$.

To this end, we have to quantify the amount of information about $f(x) = y_k$ obtained from querying $g(x) = z_i$ by calculating the pointwise mutual information as follows[3]

$$
\begin{aligned}
\mathrm{I}(f = y_k; g = z_i) &= \log \frac{\Pr(f = y_k, g = z_i)}{\Pr(f = y_k)\Pr(g = z_i)} \\
&= \log \frac{\Pr(f = y_k \mid g = z_i)}{\Pr(f = y_k)} = \log \frac{\Pr(g = z_i \mid f = y_k)}{\Pr(g = z_i)}.
\end{aligned}
\tag{1}
$$

As a consequence, the **least** number of queries needed by solving $f(x) = y_k$ will be

$$
time(f = y_k) = \frac{\mathrm{I}(f = y_k)}{\mathrm{I}(f = y_k; g = z_i)} = \frac{\log \dfrac{1}{\Pr(f = y_k)}}{\log \dfrac{\Pr(g = z_i \mid f = y_k)}{\Pr(g = z_i)}}.
\tag{2}
$$

Specifically, if $\Pr(g = z_i \mid f = y_k) = 1$, then we have

$$
time(f = y_k) = \frac{\mathrm{I}(f = y_k)}{\mathrm{I}(f = y_k; g = z_i)} = \frac{\log \dfrac{1}{\Pr(f = y_k)}}{\log \dfrac{1}{\Pr(g = z_i)}} = \frac{\mathrm{I}(f = y_k)}{\mathrm{I}(g = z_i)}.
\tag{3}
$$

It is worth noting that this result is the same as that of Shannon [1949]. In view of this, our information-based estimation of complexity naturally generalizes Shannon's theory.

Further, if the search problem $f(x) = y_k$ is Turing reducible to another function $g(x)$, then the **least** number of queries required to solve $f(x) = y_k$ will be $time(f = y_k) = \dfrac{\mathrm{I}(f = y_k)}{\mathrm{I}(f = y_k; g)}$, where $\mathrm{I}(f = y_k; g)$ is the mutual information between the event $f(x) = y_k$ and the function $g(x)$, that is,

---

[3] In theory, the pointwise mutual information can take either positive or negative values. For our present purposes, however, we assume that $\mathrm{I}(f = y_k; g = z_i) \geq 0$ for any Turing reduction $f = y_k \leq_{\mathrm{T}} g = z_i$. In such a case we have $\Pr(f = y_k \mid g = z_i) \geq \Pr(f = y_k)$ and $\Pr(g = z_i \mid f = y_k) \geq \Pr(g = z_i)$ simultaneously.

$$\mathrm{I}(f=y_k;g) = \sum_{i=1}^{m} \mathrm{Pr}(g=z_i \mid f=y_k)\mathrm{I}(f=y_k;g=z_i)$$

$$= \sum_{i=1}^{m} \mathrm{Pr}(g=z_i \mid f=y_k)\log\frac{\mathrm{Pr}(g=z_i \mid f=y_k)}{\mathrm{Pr}(g=z_i)} \qquad (4)$$

However, due to lack of information about the corresponding conditional probabilities $\mathrm{Pr}(g=z_i \mid f=y_k)$, it seems difficult to analysis the **least** number of queries theoretically.

To avoid this logical difficulty, we shall consider the **average** number of queries instead. To this end, rewrite $time(f=y_k)$ as follows

$$time(f=y_k) = \frac{\mathrm{I}(f=y_k)}{\mathrm{I}(f=y_k;g)} = \frac{\mathrm{I}(f=y_k)}{\mathrm{I}(f;g)} \times \frac{\mathrm{I}(f;g)}{\mathrm{I}(f=y_k;g)}. \qquad (5)$$

Since $\mathrm{I}(f;g)$ is the average amount of information between $f(x)$ and $g(x)$, $\dfrac{\mathrm{I}(f=y_k)}{\mathrm{I}(f;g)}$ amounts to the **average** number of queries required to solve $f(x)=y_k$ when given access to $g(x)$. As we shall see, this inherent connection places an important restriction on both lower and upper bound on the time complexity of solving $f(x)=y_k$.

With this purpose, consider the following identity

$$\frac{\mathrm{I}(f=y_k)}{\mathrm{I}(f;g)} = \frac{\mathrm{I}(f=y_k)}{\mathrm{H}(f)} \times \frac{\mathrm{H}(f)}{\mathrm{I}(f;g)}. \qquad (6)$$

This identity is natural in that $\dfrac{\mathrm{I}(f=y_k)}{\mathrm{H}(f)}$ is precisely the ratio of the amount of self-information about the event $\{f=y_k\}$ to the average amount of information contained in $f$. Also note that $\dfrac{\mathrm{H}(f)}{\mathrm{I}(f;g)}$ amounts to the **expected** number of queries required to invert $f(x)$ in case $f \leq_{\mathrm{T}} g$. Since $\mathrm{I}(f;g) \leq \min(\mathrm{H}(f),\mathrm{H}(g))$, we always have $\dfrac{\mathrm{H}(f)}{\mathrm{I}(f;g)} \geq 1$.

In fact, it turns out that $\dfrac{\mathrm{I}(f=y_k)}{\mathrm{H}(f)}$ is a **lower bound** on the query complexity of solving $f(x)=y_k$ among all possible Turing reductions. To see this, we shall consider the efficiency of an *ideal* Turing reduction, in a way similar to Carnot's work on measuring the efficiency of an ideal heat engine in thermodynamics (see Feynman *et al.* [2013]). It is well known that the most efficient heat engine is an idealized engine in which all the processes are reversible. For any reversible engine that works between temperatures $T_1$ and $T_2$, the heat $Q_1$ absorbed at $T_1$ and the heat $Q_2$ delivered at $T_2$ must be related by thermodynamics entropy, that is,

$$\frac{Q_1}{T_1} = \frac{Q_2}{T_2} = entropy.$$   (7)

*No heat engine working between temperatures $T_1$ and $T_2$ can do more work than a reversible engine*. This principle reflects that the behavior of the nature must be limited in a certain way.

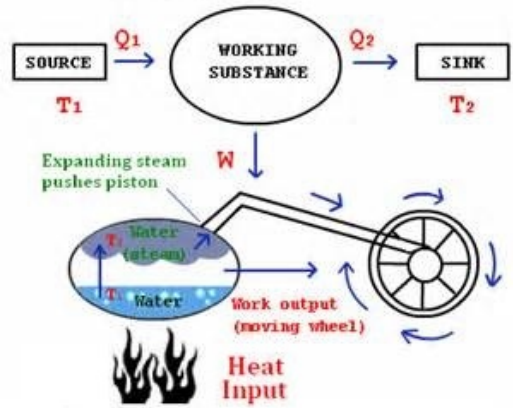

Fig. 3. Heat Engine.

The same argument applies well to reversible Turing reductions. To see this, consider a reversible Turing reduction $f = y_k \leq_T g = z_i$, that is, we assume $g = z_i$ is also reducible to $f = y_k$. In this case we say the search problem $f = y_k$ is *Turing equivalent* to $g = z_i$, denoted as $f = y_k \equiv_T g = z_i$. As a consequence, the **least** number of queries needed by solving $g(x) = z_i$ is equal to $time(g = z_i) = \dfrac{\mathrm{I}(g = z_i)}{\mathrm{I}(g = z_i; f = y_k)}$. By the symmetry of the pointwise mutual information, we obtain the following relationship

$$\frac{\mathrm{I}(f = y_k)}{time(f = y_k)} = \mathrm{I}(f = y_k; g = z_i) = \mathrm{I}(g = z_i; f = y_k) = \frac{\mathrm{I}(g = z_i)}{time(g = z_i)}.$$   (8)

Specially, if $f$ is *Turing equivalent* to $g$, then the **averag**e number of queries needed to solve $f = y_k$ and $g = z_i$ must be related by the average mutual information

$$\frac{\mathrm{I}(f = y_k)}{time(f = y_k)} = \frac{\mathrm{I}(g = z_i)}{time(g = z_i)} = \mathrm{I}(f; g).$$   (9)

According to the data processing inequality (see Thomas M. Cover, Joy A. Thomas [2006]), $f \equiv_T g$ means that the average mutual information $\mathrm{I}(f; g)$ is **maximal** among all possible Turing reductions from $f(x)$ to other functions, because $f(x)$ and $g(x)$ contain all useful information about each other.

In the ideal case, if $\mathrm{I}(f; g) = \mathrm{H}(f)$, then we obtain a perfect analogy of an ideal heat engine as follows

$$\frac{\mathrm{I}(f = y_k)}{time(f = y_k)} = \frac{\mathrm{I}(g = z_i)}{time(g = z_i)} = \mathrm{H}(f). \tag{10}$$

In this ideal case, $time(f = y_k) = \dfrac{\mathrm{I}(f = y_k)}{\mathrm{H}(f)}$ is **minimal** among all possible Turing reductions from $f(x) = y_k$ to other functions. *No algorithm can solve $f(x) = y_k$ with query complexity strictly less than* $\dfrac{\mathrm{I}(f = y_k)}{\mathrm{H}(f)}$. Once again, this limitation is the property of the nature, not a property of a particular problem.

In conclusion, working always with ideal Turing reductions, the query complexity of solving $f(x) = y_k$ *does not depend on the design of the Turing machine*. This is a perfect analogy of Carnot's brilliant conclusion: that the efficiency of a reversible heat engine *does not depend on the design of the heat engine* (see Feynman *et al.* [2013]).

Developing this spirit further, if $\dfrac{\mathrm{I}(f = y_k)}{\mathrm{H}(f)}$ is *exponential* in the size of input, then the search problem $f(x) = y_k$ can not be reduced to any other function within polynomial time. According to the Church–Turing Thesis, in this case $f(x) = y_k$ is less likely to have polynomial-time algorithm on any ordinary Turing machine. This important result can be accurately stated as follows:

**THEOREM**. *Let $f : X \to \Re$ be a simple function with values $y_1, y_2, \cdots, y_n$ for sufficiently large $n$. Given $k$, if $\dfrac{\mathrm{I}(f = y_k)}{\mathrm{H}(f)}$ is exponential in the size of input, then the probability of $f(x) = y_k$ having polynomial-time algorithm is a negative exponential in the size of input.*

This main theorem will be rigorously proved in section 4.1 on the basis of Markov's inequality. Unfortunately, the assumption of large $n$ means that this theorem does not apply to decision problems. But this difficulty is not essential, since we can consider search problems instead and work within the class of function problems (see Rich [2007], section 28.10). In fact, this theorem enables us to reduce the proof of $FP \neq FNP$ to the computation of the entropy of certain simple functions, such as the subset sum function (section 4.1). As a result, $P \neq NP$ is provable, answering a long-standing open problem (see S. Aaronson [2003]).

As it turns out, our information-based complexity estimation is a natural setting in which to study the power of randomized or probabilistic algorithms. In fact, our information-based estimation of the lower bound on repetitions of polynomial-time probabilistic algorithm is superior to that based on the Chernoff bound (see Papadimitriou [1994]).

Given a Turing reduction from $f(x) = y_k$ to $g(x)$, in theory $\dfrac{\mathrm{I}(f = y_k)}{\mathrm{I}(f;g)}$ just gives rise to an **upper bound** on the query complexity of solving $f(x) = y_k$. In the idea case, if $\dfrac{\mathrm{I}(f = y_k)}{\mathrm{I}(f;g)}$ is polynomial in the size of input and the function $g(x)$ can be computed in polynomial time, then the problem $f(x) = y_k$ also has polynomial-time algorithm. Applying to decision problems, our result provides strong evidence that $P = RP = BPP$.

Finally, using brute-force search as a benchmark for efficiency (see Kleinberg and Tardos [2005]), our results also support that $P \neq PP$. This result is obtained by estimating the lower bound on the average mutual information for any reduction-based algorithm that is more efficient than brute-force search. However, the existence of such a lower bound contradicts with the definition of the class $PP$ as desired.

The rest of this paper proceeds as follows. Section 2 reviews some definitions and lemmas in Shannon information theory and measure theory. In Section 3 we discuss how to calculate the entropy of a simple function $f : X \to \Re$ with finite values $y_1, y_2, \cdots, y_n$ and the average mutual information when the search problem $f(x) = y_k$ is reducible to other function $g(x)$. Section 4 we investigate the inherent connection between the amount of information contained in the function $f : X \to \Re$ and the time complexity of solving $f(x) = y_k$. Applying to decision problems, our results support $P = RP = BPP$ and $P \neq PP$. Section 5 focuses on several open problems worthy of further studies. Section 6 concludes the paper.

## 2. MEASURABLE FUNCTIONS

We begin with some definitions and lemmas. Those who are familiar with Shannon information theory (see Thomas M. Cover, Joy A. Thomas [2006]) and measure theory (see Patrick Billingsley [1995]) can skip directly to section 3.

### 2.1 Random Variables

A random variable is a measurable function $\xi : X \to \Re$ defined on the sample space $X$, endowed with probability measure, that is, $\mu(X) = 1$.

Generally, randomness means uncertainty. In theory, the degree of uncertainty about given random variable may be characterized by its entropy (see Shannon [1948]). Shannon entropy plays a central role in information theory as measure of information, choice, and uncertainty. Since Shannon entropy measures the average amount of information contained in given random variable, it provides an absolute limit on the best possible **least** number of choices of specifying a value of given random variable, assuming that each choice requires one unit of information.

In his 1948 paper "A Mathematical Theory of Communication", Shannon classified random variables into the discrete case and the continuous case. If $\xi : X \to \Re$ is a discrete random variable with probabilities $p_1, p_2, \cdots, p_n$, then its entropy is defined to be

$$\mathrm{H}(\xi) = \mathrm{H}(p_1, p_2, \cdots, p_n) = -\sum_{k=1}^{n} p_k \log p_k \,. \tag{11}$$

Note that the choice of a logarithmic base corresponds to the choice of a unit for measuring information. Typically, the base is taken to be $2$ and will be dropped if the context makes it clear.

As for the continuous case, to a considerable extent it can be obtained through a limiting process from the discrete case by dividing its range into a large but finite number of small regions and calculating the entropy involved on a discrete basis. As the size of the regions is decreased the entropy in general approaches as limits the proper entropy for the continuous case. However, the theory of entropy can be formulated in a completely axiomatic and rigorous manner which includes both the discrete and continuous cases. Formally, given a random variable $\xi : X \to \Re$ with probability density function $p(x)$, its Shannon entropy $\mathrm{H}(\xi)$ can be defined by means of Riemann-Stieltjes integral

$$\mathrm{H}(\xi) = -\int_{-\infty}^{+\infty} p(x) \log p(x) d\mu(x), \tag{12}$$

where $\mu$ is the Lebesgue measure if $\xi$ is continuous, and $\mu$ is the counting measure in case $\xi$ is discrete.

However, since arbitrary function may not be integrable in the sense of Riemann-Stieltjes, we have to define entropy for measurable functions on the basis of Lebesgue integral.

### 2.2 Measurable Functions

Let $(X, \mu)$ be a measurable space. A real-valued function $f : X \to \Re$ is said to be Lebesgue measurable in case $\{x \in X \mid f(x) > c\}$ is measurable for all $c \in \Re$.

If $\mu(X) = 1$, then a measurable function $f : X \to \Re$ is by definition a random variable. In this case, $X$ amounts to the sample space.

It is well known that every Lebesgue measurable function is nearly continuous, as being stated by the famous Lusin's theorem (see Patrick Billingsley 1995). On the other hand, most functions of interest in combinatorial optimization and cryptography are discrete. In order to build "a detailed, highly mathematical and more specifically analytical, theory of automata and of information ", we shall unify both the discrete and continuous cases by invoking Simple Functions (see Patrick Billingsley [1995]).

**2.3   Simple Functions**

A simple function is a real-valued function that attains only a finite number ( $n$ ) of values. Formally, a simple function is a finite linear combination of characteristic functions of measurable sets. More precisely, let $(X, \mu)$ be a measurable space, $X_1, X_2, \cdots, X_n$ be a finite partition of $X$ , and $c_1, c_2, \cdots, c_n$ be a sequence of real numbers. A simple function is a function $f : X \to \Re$ of the form

$$f(x) = \sum_{k=1}^{n} c_k \chi_{X_k}(x) \tag{13}$$

where $\chi_{X_k}$ is the characteristic function of the set $X_k$ .

It turns out that any measurable function can be approximated by simple functions. More precisely, we have the following lemma (see Patrick Billingsley 1995).

**LEMMA 2.1**.   *Let $f : X \to \Re$ be any measurable function. Then there exists a sequence of simple functions $f_n : X \to \Re$ such that $\lim_{n \to \infty} f_n(x) = f(x)$ for all $x \in X$ .*

**PROOF**.   Without loss of generality, assume $f : X \to \Re$ to be non-negative. For each $n \in \mathrm{N}$ , subdivide the range of $f : X \to [0, +\infty)$ into $2^{2n} + 1$ intervals: $[\frac{k-1}{2^n}, \frac{k}{2^n})$ for $k = 1, 2, \cdots, 2^{2n}$ and $[2^n, +\infty)$ . For each $n \in \mathrm{N}$ , define the measurable sets

$$X_{n,k} = f^{-1}([\frac{k-1}{2^n}, \frac{k}{2^n})), \forall k = 1, 2, \cdots, 2^{2n} \ and \ X_{n, 2^{2n}+1} = [2^n, +\infty) \ . \tag{14}$$

Consequently, define the sequence of increasing simple functions

$$f_n = \sum_{k=1}^{2^{2n}+1} \frac{k-1}{2^n} \chi_{X_{n,k}} \ . \tag{15}$$

Then it is easy to see that $\lim_{n \to \infty} f_n(x) = f(x)$ for all $x \in X$ (see Fig. 1). □

## 3.   INFORMATION MEASURE

Following the technical route of Shannon (1948), we shall measure the average amount of information contained in a measurable function by means of entropy.

**3.1   Entropy**

First of all, there is a natural way to obtain random variables from given measurable function. The key idea is that each measurable function can be approximated by a simple function with finite number of values (see Patrick Billingsley [1995]). As a result, these values, coupled with the probability measure of their preimages, forms a random variable in itself. So, it is natural to approximate the entropy of the given function by the entropy of the resulting random variable.

Formally, let $(X, \mu)$ be a measurable space with measure $\mu(X) < \infty$ and $f : X \to \Re$ be a measurable function. To define the entropy for $f : X \to \Re$ conducts the following steps.

(1) Approximate the given function $f : X \to \Re$ by a simple function $f_n : X \to \Re$ with values $y_1, y_2, \cdots, y_n$, by partitioning the range of $f(x)$ in the same way as computing its Lebesgue integral (see the proof of Lemma 2.1).

(2) For each value $y_k$ $(k = 1, 2, \cdots, n)$,

  (a) Define the measurable set $X_k = f_n^{-1}(y_k) = \{x \in X \mid f_n(x) = y_k\}$.

  (b) Calculate $p_k = \Pr(f = y_k) = \dfrac{\mu(X_k)}{\mu(X)}$, which stands for the probability that the output of a randomly chosen input $x$ takes the value $y_k$.

(3) Compute the sum $\mathrm{H}(f_n) = -\sum_{k=1}^{n} p_k \log p_k$ as the entropy of $f_n$.

(4) Define the entropy of $f$ to be the limit $\mathrm{H}(f) := \lim_{n \to \infty} \mathrm{H}(f_n)$, if necessary. [4]

If $\mu(X) = 1$, then a measurable function $f : X \to \Re$ is by definition a random variable. In this case, our definition of entropy coincides with that of Shannon [1948].

As a natural generalization of the Shannon entropy of random variables, the entropy of measurable functions posses well-behaved properties. To simplify notations, in the following we assume that $f : X \to \Re$ itself is a simple function taking values $y_1, y_2, \cdots, y_n$ with probabilities $p_1, p_2, \cdots, p_n$, respectively, and denote its entropy by $\mathrm{H}(f) = H(p_1, p_2, \cdots, p_n)$.

**MAXIMUM.** $\mathrm{H}(f) = \mathrm{H}(p_1, p_2, \cdots, p_n) \le \log n$, *with equality if and only if all the values are equally likely, i.e.,* $p_k = \dfrac{1}{n}$ *for all* $k = 1, 2, \cdots, n$.

---

[4] If $f(x)$ happens to be continuous, then Differential Entropy should be considered instead (see Shannon [1948]; Thomas M. Cover, Joy A. Thomas [2006]).

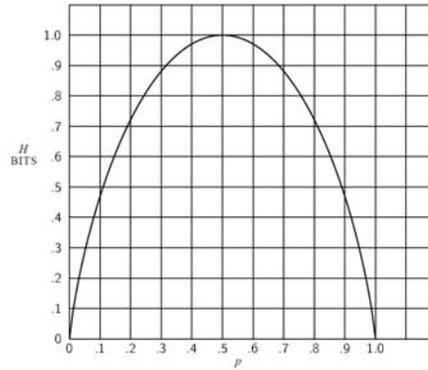

Fig. 4. Entropy in the case of two possibilities.

*Source:* Shannon, Bell System Technical Journal, vol. 27, 1948.

**CONTINUITY.** $\lim_{\varepsilon \to 0} \mathrm{H}(p_1, p_2, \cdots, p_n - \varepsilon, \varepsilon) = \mathrm{H}(p_1, p_2, \cdots, p_n)$ . *The entropy is continuous, so that changing the values of the probabilities by a very small amount should only change the entropy by a small amount. Specially, adding or removing a value with probability zero does not contribute to the entropy, that is,* $H(p_1, p_2, \cdots, p_n, 0) = H(p_1, p_2, \cdots, p_n)$ .

*Example 3.1 (Dirichlet function).* The Dirichlet function is the characteristic function of the set of rational numbers ($Q$). Formally, the Dirichlet function is defined by

$$\chi_Q(x) = \begin{cases} 1, & if \quad x \in Q \\ 0, & if \quad x \notin Q \end{cases} \tag{16}$$

The range of Dirichlet function contains only two values $0$ and $1$. It is easy to see $X_1 = \chi_Q^{-1}(1) = Q$ . Since the Lebesgue measure of the set of rational numbers is zero, we have $\mu(X_1) = \mu(Q) = 0$ . Consequently, $p_1 = 0$ , and hence $p_1 = 1 - p_0 = 1$ . Thus the entropy of the Dirichlet function is

$$\mathrm{H}(\chi_Q) = -(p_0 \log p_0 + p_1 \log p_1) = 0 \tag{17} \square$$

*Example 3.2 (Modular arithmetic).* Modular arithmetic is one of the foundations of number theory, and is widely used in cryptography. The modular function $MOD : Z \to Z_n$ is defined by $MOD(x) = x(\mathrm{mod}\, n)$ , where the modulus $n$ is a positive integer number.

(1) The image of $MOD : Z \to Z_n$ take integers between $0$ and $n-1$ .

(2) For each integer $0 \le k \le n-1$ , define the set $X_k = \{x \in Z \mid x \equiv k(\mathrm{mod}\, n)\}$ . Since every integer belongs to one and only one residue class modulo $n$ , the corresponding probability of a random input to fall into $X_k$ is $p_k = \dfrac{1}{n}$ .

(3) The entropy of the modular arithmetic is

$$\mathrm{H}(MOD) = -\sum_{k=0}^{n-1} p_k \log p_k = \log n \qquad (18)$$

As a special case, if $n = 2$, then $\mathrm{H}(MOD) = 1$. $\square$

It is worth emphasis that the entropy of a function is an attribute of this function. As such, the value of the entropy conveys important information about given function. On the other hand, the computation of the entropy may also need some important information about this function. The following two examples will illustrate this.

Firstly, we shall show that calculating the entropy of a Boolean function must be at least as difficult as the Boolean satisfiability problem (SAT). Recall that each propositional logic formula can be expressed as a Boolean function (see Shannon [1938], [1949b]).

*Example 3.3 (Boolean function).* Boolean functions play a fundamental role in questions of computational complexity theory. Formally, a Boolean function is a function $BOOL : \{0,1\}^n \to \{0,1\}$.

(1) The measure of its domain equals $\mu(\{0,1\}^n) = 2^n$, where $\mu$ is the counting measure. The range of a Boolean function contains only two values $0$ and $1$.

(2) For each value $k \in \{0,1\}$, define the set $X_k = \{x \in \{0,1\}^n \mid BOOL(x) = k\}$. Then the corresponding probability is $p_k = \dfrac{\mu(X_k)}{2^n}$. To be precise, $p_1$ is the probability of a random input $x \in \{0,1\}^n$ to satisfy $BOOL(x) = 1$.

(3) The entropy of a Boolean function is

$$\mathrm{H}(BOOL) = -(p_0 \log p_0 + p_1 \log p_1). \qquad (19)$$

It is worth emphasis that to calculate the probability $p_1$ we have to count the number of satisfying assignments of given Boolean formula. This means that we have to solve the counting problem $\#\mathrm{SAT}$, which must be at least as difficult as the decision problem SAT (see Goldreich [2008], Ding-Zhu Du, Ker-I Ko [2013]).

To see why, note that to solve the SAT problem, we are just asked to *decide* whether there is some input $x \in \{0,1\}^n$ satisfying $BOOL(x) = 1$. Even in the functional SAT problem, only *one* satisfying assignment is needed to be *found*, rather than finding out *all*. $\square$

Next, we shall show that computing the entropy of Euler's totient function is connected with some long-standing open problem, and hence it is impossible to accurately calculate so far.

*Example 3.4 (Euler's Totient Function).* In number theory, Euler's totient function $\varphi(x)$ counts the positive integers up to a given number $x$ that are relatively prime to $x$. Formally, $\varphi(x)$ is defined as the number of integers $k \leq x$ for which the greatest common divisor $\gcd(x,k) = 1$.

(1) Approximate the domain of $\varphi(x)$ by the set of natural numbers between $1$ and $n$. Then the range of $\varphi(x)$ consists of *totient numbers* up to $n$. So the counting measure of the rang of $\varphi(x)$ coincides with the number of totient numbers up to $n$, that is equal to $\dfrac{n}{\log n} \exp((C + o(1))(\log\log\log x)^2)$ for a constant $C = 0.8178143\cdots$(see Ford [1998]).

(2) For each totient number $k$ between $1$ and $\varphi(n)$,

    (a) Define the measurable sets $X_k = \{x \le n \,|\, \varphi(x) = k\}$. It is easy to see that the counting measure of $X_k$ equals the *multiplicity* of the totient number $k$.

    (b) Calculate the corresponding probability $p_k = \dfrac{\mu(X_k)}{n}$, which is the probability of a random integer $x \le n$ to satisfy $\varphi(x) = k$.

The difficulty in computing the entropy of Euler's totient function lies in the fact that the behavior of the multiplicity of totient numbers is not clear (see Ford [1999]). Indeed, there is a famous unsolved problems concerning the multiplicity of totient numbers, namely, Carmichael's Conjecture. In 1907, Carmichael Conjectured that for every $k$, the equation $\varphi(x) = k$ has either no solutions or at least two solutions. In other words, no totient can have multiplicity 1. In our notations, Carmichael's Conjecture asserts $\mu(X_k) \ge 2$ for all totient number $k$. Since Carmichael's Conjecture remains an open problem (see Ford [1999]), so far it is impossible to accurately calculate the entropy of Euler's totient function. □

## 3.2  Information Amount

The entropy of a function characterizes the uncertainty about the image $f(x)$ of *random* input $x$. In other words, the entropy of a function $f : X \to \Re$ is the average amount of information needed for specifying the value of a *random* input $x \in X$.

But, given a measurable function $f : X \to \Re$, in complexity theory we are just interested with the amount of information associated with specifying some particular values. For example, to decode a Public-key cryptography, we have to invert the enciphering function $y = f(x)$ given a cipher text $y_k$. Obviously, the given cipher text $y_k$ will convey some information about the corresponding preimage(s). Indeed, given the enciphering function $y = f(x)$ and arbitrary cipher text $y_k$, the goal of cryptanalyst is to gain as much information as possible about the plaintext $x \in f^{-1}(y_k)$. However, for cryptographic purposes, it may be possible that one can decrypt *one* arbitrary cipher text $y_k$ without inverting the enciphering function $f$ directly. Intuitively, decrypting *one* cipher text $y_k$ may not be computationally equivalent to inverting the enciphering function $f$, which means decrypting *all* cipher texts (see Diffie and Hellman [1976]). To overcome this logical difficulty, we have to make good use of the information conveyed by the given value $y_k$.

Formally, let $(X, \mu)$ be a measurable space with measure $\mu(X) < \infty$ and $f : X \to \Re$ be a simple function with values $y_1, y_2, \cdots, y_n$. Then the self-information associated with $\{f = y_k\}$ is equal to $\mathrm{I}(f = y_k) = -\log p_k$, where $p_k = \mathrm{Pr}(f = y_k) = \dfrac{\mu(f^{-1}(y_k))}{\mu(X)}$. The total amount of information contained in the function $f : X \to \Re$ with values $y_1, y_2, \cdots, y_n$ is equivalent to the sum $\mathrm{I}(f) = \sum_{k=1}^{n} \mathrm{I}(f = y_k) = -\sum_{k=1}^{n} \log p_k$.

*Example 3.5 (Discrete Logarithm).* The discrete logarithm requires inverting a modular exponentiation function. Formally, a modular exponentiation function is $EXP(x) = b^x (\mathrm{mod}\, n)$, where the base $b \geq 2$ is a primitive root modulo $n$. In other words, $b$ is a generator of the multiplicative group of integers modulo $n$.

(1) Since $b$ is a generator of the multiplicative group of integers modulo $n$, $b^x(\mathrm{mod}\, n)$ ranges from $0$ to $n-1$, and hence the measure of the range equals $n$.

(2) For value $k$ between $0$ and $n-1$, define the set $X_k = \{x \,|\, b^x \equiv k(\mathrm{mod}\, n)\}$. Note that $b$ is a primitive root modulo $n$ implies that $X_k$ contains only one element for each value $k$ (see Hua 1982). Consequently, the probability of a random input to satisfy $b^x \equiv k(\mathrm{mod}\, n)$ is $p_k = \dfrac{1}{n}$.

(3) The entropy of modular exponentiation function is

$$\mathrm{H}(EXP) = -\sum_{k=0}^{n-1} p_k \log p_k = \log n. \tag{20}$$

(4) To compute discrete logarithm, we are required to find out $x < n$ such that $b^x \equiv k(\mathrm{mod}\, n)$. The self-information associated with the event $\{EXP = k\}$ is $\mathrm{I}(EXP = k) = -\log p_k = \log n = \mathrm{H}(EXP)$. $\square$

*Example 3.6 (Rabin Function).* The Rabin function is defined by squaring modulo $n = q_1 q_2$, where $q_1$ and $q_2$ are distinct odd prime numbers. Formally, $RABIN(x) = x^2(\mathrm{mod}\, n)$. It is well known that inverting the Rabin function is computationally equivalent to integer factorization in the sense of polynomial-time reduction (see Rabin [1979]).

(1) It suffices to restrict the domain of $RABIN(x)$ to a reduced residue system modulo $n$. So the counting measure of the domain of $RABIN(x)$ equals $\varphi(n) = (q_1 - 1)(q_2 - 1)$, where $\varphi(\cdot)$ is the Euler's totient function. The range of $RABIN(x)$ consists of all quadratic residue modulo $n$, and hence the measure of the range equals $\dfrac{\varphi(n)}{4}$ (see Hua [1982]).

(2) For each quadratic residue $y_k$, define the set $X_k = \{x \mid x^2 \equiv y_k \,(\mathrm{mod}\, n)\}$. Since $n = q_1 q_2$ is a semiprime, $X_k$ exactly has $4$ elements (see Hua [1982]). Consequently, the corresponding probability is $p_k = \dfrac{4}{\varphi(n)}$.

(3) The entropy of Rabin function is

$$\mathrm{H}(RABIN) = -\sum_{k=0}^{\frac{\varphi(n)}{4}} p_k \log p_k = \log \frac{\varphi(n)}{4}. \tag{21}$$

(4) To decode the Rabin cryptosystem, we are required to find out $x < n$ such that $x^2 \equiv y_k \,(\mathrm{mod}\, n)$. The self-information associated with the event $\{RABIN = y_k\}$ is $\mathrm{I}(RABIN = y_k) = -\log p_k = \log \dfrac{\varphi(n)}{4} = \mathrm{H}(RABIN)$. $\square$

*Example 3.7 (RSA function).* Formally, the *RSA* function is defined to be $RSA(x) = x^e \,(\mathrm{mod}\, n)$, where $n = q_1 q_2$, the product of distinct odd prime numbers, and $e \geq 3$ is relatively prime to $\varphi(n) = (q_1 - 1)(q_2 - 1)$ (see Rivest, Shamir, and Adleman [1978]).

(1) It suffices to restrict the domain of $RSA(x)$ to a reduced residue system modulo $n$. So the counting measure of the domain of $RSA(x)$ equals $\varphi(n) = (q_1 - 1)(q_2 - 1)$. The range of $RSA(x)$ consists of all $e$-th power residue modulo $n$, and hence the measure of the range equals $\varphi(n)$.

(2) For each $e$-th residue $y_k$, define the set $X_k = \{x < n \mid x^e \equiv y_k \,(\mathrm{mod}\, n)\}$. Since $e$ is relatively prime to $\varphi(n)$, the congruence equation $x^e \equiv k \,(\mathrm{mod}\, n)$ has exactly $1$ solution. Consequently, the corresponding probability is $p_k = \dfrac{1}{\varphi(n)}$.

(3) The entropy of *RSA* function is

$$\mathrm{H}(RSA) = -\sum_{k=0}^{\varphi(n)} p_k \log p_k = \log \varphi(n). \tag{22}$$

(4) To decode the *RSA* cryptosystem, we are required to find out $x < n$ such that $x^e \equiv y_k \,(\mathrm{mod}\, n)$. The self-information associated with the event $\{RSA = y_k\}$ is $\mathrm{I}(RSA = y_k) = -\log p_k = \log \varphi(n) = \mathrm{H}(RSA)$.

But, the *RSA* cryptosystem would be broken if the number $n$ could be factored or if $\varphi(n)$ could be computed without factoring $n$. Indeed, the *RSA* function is widely conjectured to be trapdoor one-way. The trapdoor is the information about $d \equiv e^{-1}(\mathrm{mod}\,\varphi(n))$, the multiplicative inverse of $e$ modulo $\varphi(n)$. In fact, the prime factors $q_1, q_2$ and the totient number $\varphi(n)$ are also part of the trapdoor information because they can be used to calculate $d$. If the factorization $n = q_1 q_2$ is known, Euler's totient function $\varphi(n) = (q_1 - 1)(q_2 - 1)$ can be computed, then we can solve for $d$ given $ed \equiv 1(\mathrm{mod}\,\varphi(n))$. Using this trapdoor information, we can easily recover $x$ from $y_k = x^e(\mathrm{mod}\,n)$ via computing $x = y_k^d(\mathrm{mod}\,n)$. □

*Example 3.8 (Integer Factorization).* The integer factorization problem requires inverting the integer multiplication function. Formally, the integer multiplication function is a binary operation on the set of natural numbers defined by $M(u,v) = uv$, where $u, v \in \mathrm{N}$ are natural numbers.

(1) Approximate the range of $M(u,v)$ by the set of natural numbers between $1$ and $n$.

(2) For each value $k$ between $1$ and $n$,

    (a) Define the measurable sets $X_k = \{(u,v) \in \mathrm{N} \times \mathrm{N} \mid uv = k\}$. It is easy to see that the counting measure of $X_k$ equals the number of ways that the integer $k$ can be written as a product of two integers. In number theory, the number of divisors of an integer $k$ is usually denoted as $d(k)$, the divisor function. That is, $\mu(X_k) = d(k)$. Summing over all divisor function, we get the divisor summatory function $D(n) = \sum_{k=1}^{n} d(k)$. In big-O notation, Dirichlet demonstrated that $D(n) = n \log n + n(2\gamma - 1) + \mathrm{O}(\sqrt{n})$, where $\gamma = 0.5772\cdots$ is the Euler's gamma constant. Improving the bound $\mathrm{O}(\sqrt{n})$ in this formula is known as Dirichlet's Divisor Problem (see Hua [1982]).

    (b) Calculate the corresponding probability $p_k = \dfrac{\mu(X_k)}{\mu(X)} = \dfrac{d(k)}{D(n)}$, which is the probability of a random integer pair $(u,v)$ below the hyperbola $uv = n$ to satisfy $uv = k$.

(3) The self-information associated with factoring $n$ is
$$\mathrm{I}(M = n) = -\log p_n = \log \frac{D(n)}{d(n)} \approx \log \frac{n \log n}{d(n)}.$$

The difficulty in estimating the self-information associated with factoring $n$ is that the behavior of the divisor function $d(n)$ is irregular (see Hua Lo-keng [1982]). However, a lower bound can be obtained by invoking the following property of the divisor function: for all $\varepsilon > 0$, the divisor function satisfies the inequality $d(n) < n^{\varepsilon}$ (see Hua [1982]). As a result, a lower bound on the self-information associated with factoring $n$ is $\log(n^{1-\varepsilon}\log n)$. In the extreme case, if $n$ is a semiprime with only two prime factors, then $d(n) = 4$ and the self-information associated with factoring $n$ is $\mathrm{I}(M = n) \approx \log\dfrac{n\log n}{4}$. $\square$

### 3.3 Conditional Entropy

To solve the search problem $f(x) = y_k$ we have to make good use of the information hidden in the function $f : X \rightarrow \Re$ and the information conveyed by the given value $y_k$.

To illustrate, consider the famous subset sum problem, which can be stated as follows: given a set of non-zero integers $a_1, a_2, \cdots, a_n \in Z$, is there a non-empty subset that adds up to $0$ (see Karp [1972])? The corresponding subset sum function is given by

$$SUM(x) = a_1 x_1 + a_2 x_2 + \cdots + a_n x_n, x_i \in \{0,1\}, i = 1,2,\cdots,n. \tag{23}$$

In terms of the subset sum function, the subset sum problem amounts to the existence of some $x \in \{0,1\}^n$ such that $SUM(x) = 0$. But if $a_1 + a_2 + \cdots + a_n \neq 0$, then the assumption of $SUM(x) = 0$ implies that the components of $x$ can not be *all* positive. As such, the search space may be dramatically narrowed down. However, due to lack of information about the distribution of the conditional probabilities under the condition $SUM = 0$, it seems difficult to analysis the conditional entropy in detail theoretically (see Yao [1982] for discussions).

In theory, the information conveyed by the given value $y_k$ may enable us to reduce a search problem to the corresponding decision problem (see Rich [2007]; Goldreich [2008]).

**SEARCH PROBLEM**. Given the function relation $f : X \rightarrow \Re$ and arbitrary value $y_k \in Y$, *find* some $x \in X$ such that $f(x) = y_k$.

**DECISION PROBLEM**. Given the function relation $f : X \rightarrow \Re$ and a value $y_k \in \Re$, *decide* whether there exists $x \in X$ satisfying $f(x) = y_k$.

Decision problems are important in that most well-studied complexity classes are defined as classes of decision problems. But, this kind of reduction may not make full of the information hidden in the function $y = f(x)$.

To illustrate, consider the optimization problem of a real-valued function $y = f(x)$ on a closed interval $[a,b]$. Fermat's theorem states that if $f(x)$ is *differentiable* and has a local extremum at $x_0 \in (a,b)$, then its derivative $f^{'}(x)$ must satisfy $f^{'}(x_0) = 0$. By using Fermat's theorem, the local extremum of $f(x)$ is found by solving an equation $f'(x_0) = 0$. As a result, the information of differentiability of $f(x)$ and optimality of $f(x_0)$ enable us to reduce the optimization problem $\min\limits_{x \in (a,b)} f(x)$ to another search problem $f'(x_0) = 0$ (see Fig. 2).

In some idea cases, the information hidden in the function $f : X \to \Re$ and the information conveyed by the given value $y_k$ may enable us to reduce the search problem $f(x) = y_k$ to another search problem $g(x) = z_i$ (see Karp [1972] for examples). This means that there exists an algorithm that solves $f(x) = y_k$ by calling as a *subroutine* another algorithm that solves $g(x) = z_i$ (see Goldreich [2008]). As such, to quantify the average amount of information about $f(x)$ obtained from $g(x)$ we have to calculate the conditional entropy $\mathrm{H}(f \mid g)$ (see Shannon [1948]).

Firstly, consider the entropy of $\mathrm{H}(f \mid g = z_i)$ conditioned on $g(x) = z_i$

$$\mathrm{H}(f \mid g = z_i) = -\sum_{k=1}^{n} \mathrm{Pr}(f = y_k \mid g = z_i) \log \mathrm{Pr}(f = y_k \mid g = z_i). \qquad (24)$$

Secondly, the condition entropy $\mathrm{H}(f \mid g)$ is the weighted sum of $\mathrm{H}(f \mid g = z_i)$ for each possible value of $g(x)$

$$\begin{aligned}
\mathrm{H}(f \mid g) &= \sum_{i=1}^{m} \mathrm{Pr}(g = z_i) \mathrm{H}(f \mid g = z_i) \\
&= -\sum_{i=1}^{m} \mathrm{Pr}(g = z_i) \sum_{k=1}^{n} \mathrm{Pr}(f = y_k \mid g = z_i) \log \mathrm{Pr}(f = y_k \mid g = z_i). \\
&= -\sum_{i=1}^{m} \sum_{k=1}^{n} \mathrm{Pr}(f = y_k, g = z_i) \log \mathrm{Pr}(f = y_k \mid g = z_i)
\end{aligned} \qquad (25)$$

Note that $\mathrm{H}(f \mid g) = 0$ if and only if the value of $f$ is completely determined by the value of $g$.

*Example 3.9 (Many-one Reduction).* Viewed as a special case and stronger form of Turing reductions, many-one reductions among decision problems can be defined in terms of formal languages over the alphabet $\{0,1\}$ (see Thomas H Cormen *et al.* [2009]). Formally, a language $L_1$ is many-one reducible to $L_2$ if there is a total computable function $\tau : \{0,1\}^* \to \{0,1\}^*$ such that $w \in L_1$ if and only if $\tau(w) \in L_2$ .[5] In terms of characteristic functions, this gives rise to the following conditional probabilities

$$\begin{array}{ll} \Pr(\chi_{L_2} = 0 \mid \chi_{L_1} = 0) = 1, & \Pr(\chi_{L_2} = 1 \mid \chi_{L_1} = 0) = 0, \\ \Pr(\chi_{L_2} = 0 \mid \chi_{L_1} = 1) = 0, & \Pr(\chi_{L_2} = 1 \mid \chi_{L_1} = 1) = 1. \end{array} \quad (26)$$

Viewed as a noiseless channel, it is routine to check that $\mathrm{H}(\chi_{L_2} \mid \chi_{L_1}) = 0$ . Note that with many-one reduction solving $\chi_{L_2} = 1$ is at least *as hard as* solving $\chi_{L_1} = 1$ (see Papadimitriou [1994], chapter 8; Kleinberg and Tardos [2005], chapter 8). □

*Example 3.10 (The class RP ).* One of the important questions in complexity theory is whether randomized or probabilistic algorithms are more powerful than their deterministic counterparts (see Papadimitriou [1994]). The most famous concrete formulation of this question regards the power of the class $RP$ , the set of decision problems that have Randomized, Polynomial time algorithms (see Rich [2007], section 30.2; Papadimitriou [1994]).

For convenience, we shall define decision problems in terms of formal languages. Formally, a language $L$ over the alphabet $\{0,1\}$ belongs to $RP$ if and only if there exists some randomized Turing machine that runs in polynomial time for deciding $L$ such that: if $w \notin L$ then all computation halts with rejection, if $w \in L$ then $M$ accepts $w$ with probability $1 - \varepsilon$ for arbitrarily small constant $0 < \varepsilon < \frac{1}{2}$ . It is worth emphasis that $\varepsilon$ must be independent of the input to the algorithm.

It is clear that the class $RP$ lies somewhere between $P$ and $NP$ (see Papadimitriou [1994]). One of the most famous problems that was known to be in $RP$ is the problem of determining whether a given number is (not) a prime number. However, Agrawal, Kayal and Saxena [2004] have shown that Primality test turns out to be in $P$ (see section 4.2 for details). Based on the hardness-randomness tradeoffs (see Yao [1982]), it is widely conjectured, but unproven, that $RP = P$ (see Papadimitriou [1994]).

Traditionally, the fundamental paradigm in derandomization is to trade hardness for randomness (see Yao [1982]). However, since randomness means uncertainty, it turns out that our definition of entropy is a natural setting in which to study randomization. To see this, take a language $L \subseteq \{0,1\}^*$ as fixed and consider its characteristic function

---

[5] Without loss of generality, it is assumed that $\tau(L_1) = L_2$ . For our present purposes, it doesn't matter how many instances of $L_1$ reduce to one and the same instance of $L_2$ . The essence of a many-one reduction is that it needs to preserve the correctness of the answer.

$$\chi_L(w) = \begin{cases} 1, & if \quad w \in L \\ 0, & if \quad w \notin L \end{cases}. \tag{27}$$

Let $p_1$ be the probability of acceptance, and $p_0 = 1 - p_1$ be the probability of rejection. Then the entropy of the given langue $L$ equals $H(\chi_L) = -(p_0 \log p_0 + p_1 \log p_1)$.

On the other hand, the definition of the class $RP$ gives rise to a function $\chi_L^\varepsilon$ from $\{0,1\}^*$ to the set $\{0,1\}$, with probabilities $\Pr(\chi_L^\varepsilon = 1) = p_1(1-\varepsilon)$ and $\Pr(\chi_L^\varepsilon = 0) = p_0 + p_1\varepsilon$. So the entropy of $\chi_L^\varepsilon$ is

$$H(\chi_L^\varepsilon) = -[(p_0 + p_1\varepsilon)\log(p_0 + p_1\varepsilon) + p_1(1-\varepsilon)\log p_1(1-\varepsilon)] \tag{28}$$

The language $L$ in the class $RP$ yields the following conditional probabilities

$$\begin{aligned} \Pr(\chi_L^\varepsilon = 0 \mid \chi_L = 0) = 1, \quad & \Pr(\chi_L^\varepsilon = 1 \mid \chi_L = 0) = 0, \\ \Pr(\chi_L^\varepsilon = 0 \mid \chi_L = 1) = \varepsilon, \quad & \Pr(\chi_L^\varepsilon = 1 \mid \chi_L = 1) = 1 - \varepsilon. \end{aligned} \tag{29}$$

Then the conditional entropy $H(\chi_L^\varepsilon \mid \chi_L)$ is by definition equal to

$$\begin{aligned} H(\chi_L^\varepsilon \mid \chi_L) &= \Pr(\chi_L = 0)H(\chi_L^\varepsilon \mid \chi_L = 0) + \Pr(\chi_L = 1)H(\chi_L^\varepsilon \mid \chi_L = 1) \\ &= \Pr(\chi_L = 0)\Pr(\chi_L^\varepsilon = 0 \mid \chi_L = 0)\log\frac{1}{\Pr(\chi_L^\varepsilon = 0 \mid \chi_L = 0)} \\ &+ \Pr(\chi_L = 0)\Pr(\chi_L^\varepsilon = 1 \mid \chi_L = 0)\log\frac{1}{\Pr(\chi_L^\varepsilon = 1 \mid \chi_L = 0)} \\ &+ \Pr(\chi_L = 1)\Pr(\chi_L^\varepsilon = 0 \mid \chi_L = 1)\log\frac{1}{\Pr(\chi_L^\varepsilon = 0 \mid \chi_L = 1)} \\ &+ \Pr(\chi_L = 1)\Pr(\chi_L^\varepsilon = 1 \mid \chi_L = 1)\log\frac{1}{\Pr(\chi_L^\varepsilon = 1 \mid \chi_L = 1)} \\ &= -p_1[\varepsilon\log\varepsilon + (1-\varepsilon)\log(1-\varepsilon)] \end{aligned} \tag{30}$$

It is easy to see that $\lim_{\varepsilon \to 0} H(\chi_L^\varepsilon \mid \chi_L) = 0$. $\square$

### 3.4 Mutual Information

Generally, the information hidden in the function $f : X \to \Re$ and the information conveyed by the given value $y_k$ might enable a Turing reduction from the search problem $f(x) = y_k$ to another search problem $g(x) = z_i$ (see Karp [1972] for examples). In order to quantify the amount of information about $f(x) = y_k$ obtained from querying $g(x) = z_i$ we have to calculate the pointwise mutual information as follows

$$\begin{aligned}
\mathrm{I}(f = y_k ; g = z_i) &= \log \frac{\Pr(f = y_k, g = z_i)}{\Pr(f = y_k)\Pr(g = z_i)} \\
&= \log \frac{\Pr(f = y_k \mid g = z_i)}{\Pr(f = y_k)} = \log \frac{\Pr(g = z_i \mid f = y_k)}{\Pr(g = z_i)}
\end{aligned} \tag{31}$$

Specifically, if $\Pr(g = z_i \mid f = y_k) = 1$, then we have

$$\mathrm{I}(f = y_k ; g = z_i) = \log \frac{1}{\Pr(g = z_i)} = \mathrm{I}(g = z_i). \tag{32}$$

Similarly, if $\Pr(f = y_k \mid g = z_i) = 1$, then $\mathrm{I}(f = y_k ; g = z_i) = \mathrm{I}(f = y_k)$.

*Example 3.11 (Fermat's Theorem).* Suppose that a *differentiable* function $f(x)$ on an open interval $(a, b)$ has $r$ local extrema $y_0, y_1, \cdots, y_{r-1}$. Then Fermat's theorem says that the conditional probability $\Pr(f^{'} = 0 \mid f = y_k) = 1$ for any $k = 0, 1, \cdots r - 1$. Consequently, the pointwise mutual information is equal to

$$\mathrm{I}(f = y_k ; f^{'} = 0) = \log \frac{\Pr(f^{'} = 0 \mid f = y_k)}{\Pr(f^{'} = 0)} = \log \frac{1}{\Pr(f^{'} = 0)} = \mathrm{I}(f^{'} = 0). \tag{33} \square$$

Further, if the search problem $f(x) = y_k$ is Turing reducible to another function $g(x)$, then to estimate of the complexity of solving $f(x) = y_k$ we have to measure the mutual information between the event $f(x) = y_k$ and the function $g(x)$, that is

$$\begin{aligned}
\mathrm{I}(f = y_k ; g) &= \sum_{i=1}^{m} \Pr(g = z_i \mid f = y_k)\mathrm{I}(f = y_k ; g = z_i) \\
&= \sum_{i=1}^{m} \Pr(g = z_i \mid f = y_k) \log \frac{\Pr(g = z_i \mid f = y_k)}{\Pr(g = z_i)}
\end{aligned} \tag{34}$$

Note that even though the pointwise mutual information may be negative or positive, the mutual information $\mathrm{I}(f = y_k ; g)$ is always positive, that is, $\mathrm{I}(f = y_k ; g) \geq 0$.[6] This means that, if $f(x) = y_k$ is Turing reducible to $g(x)$,

---

then we can, on average, obtain useful information about $f(x) = y_k$ from querying $g(x)$.

*Example 3.12 (Intermediate Value Theorem).* Given a *continuous* function $f(x)$ on a closed interval $[a, b]$ such that $f(a)f(b) < 0$, the intermediate value theorem asserts that there exists at least one $x_0 \in (a, b)$ satisfying $f(x_0) = 0$. The well-known bisection algorithm repeatedly generates a sequence $[a_k, b_k]$ of subintervals with $f(a_k)f(b_k) < 0$ on which a zero $x_0$ is known to lie for further processing. At the $k$th step the method divides the interval in two by computing the midpoint $\dfrac{a_k + b_k}{2}$ of the interval and deciding the sign $\text{sgn}(f(\dfrac{a_k + b_k}{2}))$ of the value of the function at that point. The method then selects the subinterval $[a_{k+1}, b_{k+1}] \subset [a_k, b_k]$ with $f(a_{k+1})f(b_{k+1}) < 0$.

In this way, the intermediate value theorem enables us to reduce the zero-finding problem $f(x) = 0$ to the sign function $\text{sgn}(f(x))$. Since $\text{sgn}(f(\dfrac{a_k + b_k}{2}))$ takes $+1$ and $-1$ with equal probability, the intermediate value theorem implies that the mutual information is equal to

$$\text{I}(f = 0; \text{sgn}(f)) = \text{H}(\text{sgn}(f)) = 1. \tag{35} \square$$

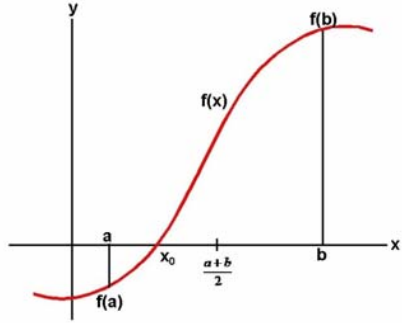

Fig. 5. The Intermediate Value Theorem.

$$-\text{I}(f = y_k; g) = \sum_{i=1}^{m} \Pr(g = z_i \mid f = y_k) \log \frac{\Pr(g = z_i)}{\Pr(g = z_i \mid f = y_k)}$$

$$\leq \log \sum_{i=1}^{m} \Pr(g = z_i \mid f = y_k) \frac{\Pr(g = z_i)}{\Pr(g = z_i \mid f = y_k)}$$

$$= \log \sum_{i=1}^{m} \Pr(g = z_i) = \log 1 = 0$$

Finally, if the function $f(x)$ is Turing reducible to another function $g(x)$, then for any $x \in X$, $f(x)$ can be computed by a Turing machine when given access to the oracle $g$ (see Goldreich [2008]). In this case, we have to measure the average mutual information between $f(x)$ and $g(x)$, which is by definition equal to the expected value of pointwise mutual information, that is

$$
\begin{aligned}
\mathrm{I}(f;g) &= \sum_{k=1}^{n} \sum_{i=1}^{m} \mathrm{Pr}(f = y_k, g = z_i) \mathrm{I}(f = y_k; g = z_i) \\
&= \sum_{k=1}^{n} \sum_{i=1}^{m} \mathrm{Pr}(f = y_k, g = z_i) \log \frac{\mathrm{Pr}(f = y_k, g = z_i)}{\mathrm{Pr}(f = y_k)\mathrm{Pr}(g = z_i)}
\end{aligned}
\tag{36}
$$

Intuitively, mutual information measures the information that $f(x)$ and $g(x)$ share. For example, if $f(x)$ and $g(x)$ take values independently, then knowing values of $g(x)$ does not give any information about $f(x)$ and vice versa, so their mutual information is zero.

The average mutual information is by definition symmetric and can be equivalently expressed as

$$
\mathrm{I}(f;g) = \mathrm{H}(f) - \mathrm{H}(f \mid g) = \mathrm{H}(g) - \mathrm{H}(g \mid f) = \mathrm{I}(g;f).
\tag{37}
$$

Further, It is easy to see that $\mathrm{I}(f;g) = \mathrm{E}(\mathrm{I}(f = y_k; g))$, where $\mathrm{E}(\cdot)$ is the mathematical expectation operator, conditioned on all possible value $y_k$ of $f(x)$. From this it follows that $\mathrm{I}(f;g) \geq 0$. In fact, it is routine to check the following inequality

$$
0 \leq \mathrm{I}(f;g) \leq \min(\mathrm{H}(f), \mathrm{H}(g)).
\tag{38}
$$

Developing this spirit further, we can get the following analogy of the famous Data Processing Inequality, which captures the fundamental nature of information theory (see Thomas M. Cover, Joy A. Thomas [2006]).

**DATA PROCESSING INEQUALITY.** *Given a sequence of Turing reductions $f \leq_{\mathrm{T}} g$ and $g \leq_{\mathrm{T}} h$, we have the following decreasing sequence of information*

$$
\mathrm{H}(f) \geq \mathrm{I}(f;g) \geq \mathrm{I}(f;h).
\tag{39}
$$

The data processing inequality implies that, if a search problem $f(x) = y$ is Turing reducible to another problem $g(x) = z$, then, on average, any Turing reduction from $g(x) = z$ to any other search problem $h(x) = w$ may decrease information about the original problem $f(x) = y$. In short, Turing reduction may lose information.

The data processing inequality is dual to the Second Law of Thermodynamics, which asserts that the entropy of the universe is always increasing (see

Feynman *et al.* [2013]). Only in reversible processes does the entropy remain constant. A reversible process is an idealization in which we have made the increase in entropy minimal.

## 4. TIME COMPLEXITY

Each algorithm uses a sequence of *elementary operations* to complete the desired work, no matter how complex the algorithm is. With regards to a Turing machine, an elementary operation can be defined as one move of its tape (see Turing [1948]). Indeed, in his 1948 essay Turing wrote: "...However, the tape can be moved back and forth through the machine, this being one of the elementary operations of the machine". The *time complexity* of an algorithm is a measure of how many times the tape moves during its computation.

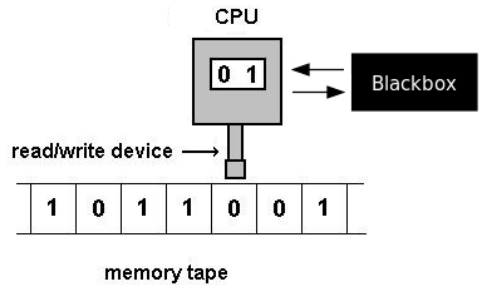

Fig. 6. Oracle Turing Machine.

On the other hand, the amount of information associated with one elementary operation is equal to the additional amount of information required to specify the state of the tape after one move of it. If we assume that one elementary operation yields one unit of information on a Turing machine, then the time complexity coincides with the total amount of information to be obtained.

This implies that there must exist an inherent connection between the amount of information contained in the function $f : X \rightarrow \Re$ and the time complexity of solving $f(x) = y_k$ (for similar ideas see Packel and Traub [1987]; Yao, A.C. [1988]; Shah and Sharma [2000] ; Tadao Takaoka, Yuji Nakagawa [2010]).

### 4.1 Query Complexity

According to the Church–Turing Thesis (see Goldreich [2008]), a Turing reduction is the most general form of an effectively calculable reduction. The notion of Turing reductions can be modeled as *oracle Turing machines* (see Goldreich [2008]; Ding-Zhu Du, Ker-I Ko [2013]), which are at least as powerful as ordinary Turing machines. Formally, the search problem $f(x) = y_k$ is Turing reducible to another search problem $g(x) = z_i$ means that there exists a Turing machine that can solve $f(x) = y_k$ when given access to an oracle for $g(x) = z_i$. During its computation, the oracle machine may query the oracle arbitrarily often with possibly different queries (see Sanjeev Arora and Boaz Barak [2009]).

Traditionally, the oracle is described as a "black box" capable of magically solving $g(x) = z_i$ in a single step (see Kleinberg and Tardos 2005, section 8.1). Due to the lack of necessary tools needed to open the "black box" of oracle

Turing machines, we have to limit the number of oracle queries and/or to limit the computational resources that the program implementing the Turing reduction may use. This directly leads to the notion of polynomial-time Turing reductions or Cook reductions (see Ding-Zhu Du, Ker-I Ko [2013]; Goldreich [2008]).

However, to estimate the time complexity of solving the search problem $f(x) = y_k$ via ordinary Turing machines we have to estimate the **least** number of queries, which in turn can be estimated by means of the amount of information associated with the function in question. In this way we shall be able to open the "black box" of oracle Turing machines.

To be precise, assume there is a *Turing reduction* $f = y_k \leq_T g = z_i$. Then there exists an algorithm that solves $f(x) = y_k$ by calling as a *subroutine* another algorithm that solves $g(x) = z_i$. Obviously, to achieve the best performance, any algorithm for solving $f(x) = y_k$ must consider the *tradeoff* between the number of queries and the complexity of each query. If each query can be done in polynomial time, then to estimate the time complexity of solving $f(x) = y_k$ we just need to estimate the **least** number of queries, denoted as $time(f = y_k)$.

Recall that the self-information associated with $\{f = y_k\}$ is equal to $I(f = y_k) = -\log p_k$, where $p_k = \Pr(f = y_k)$, and the amount of information about $f(x) = y_k$ provided by querying the solution(s) of $g(x) = z_i$ is exactly equal to the pointwise mutual information $I(f = y_k ; g = z_i)$. As a result, the **least** number of queries needed by solving $f(x) = y_k$ will be

$$time(f = y_k) = \frac{I(f = y_k)}{I(f = y_k ; g = z_i)} = \frac{\log \dfrac{1}{\Pr(f = y_k)}}{\log \dfrac{\Pr(g = z_i \mid f = y_k)}{\Pr(g = z_i)}}. \tag{40}$$

Specifically, if $\Pr(g = z_i \mid f = y_k) = 1$, then we have

$$time(f = y_k) = \frac{I(f = y_k)}{I(f = y_k ; g = z_i)} = \frac{\log \dfrac{1}{\Pr(f = y_k)}}{\log \dfrac{1}{\Pr(g = z_i)}} = \frac{I(f = y_k)}{I(g = z_i)}. \tag{41}$$

It is worth noting that this result is the same as that of Shannon [1949]. In view of this, our information-based estimation of complexity naturally generalizes Shannon's theory.

Further, if the search problem $f(x) = y_k$ is Turing reducible to $g(x)$, then the average amount of information about $f(x) = y_k$ provided by each query is equal to the mutual information $\mathrm{I}(f = y_k; g)$. As a consequence, the **least** number of queries required to solve $f(x) = y_k$ will be $time(f = y_k) = \dfrac{\mathrm{I}(f = y_k)}{\mathrm{I}(f = y_k; g)}$. However, due to lack of information about the corresponding conditional probabilities, it seems difficult to analysis the **least** number of queries theoretically.

To avoid this logical difficulty, we shall consider the **average** number of queries instead. To this end, rewrite $time(f = y_k)$ as follows

$$time(f = y_k) = \frac{\mathrm{I}(f = y_k)}{\mathrm{I}(f = y_k; g)} = \frac{\mathrm{I}(f = y_k)}{\mathrm{I}(f; g)} \times \frac{\mathrm{I}(f; g)}{\mathrm{I}(f = y_k; g)}. \tag{42}$$

Since $\mathrm{I}(f; g)$ is the average amount of information between $f(x)$ and $g(x)$, $\dfrac{\mathrm{I}(f = y_k)}{\mathrm{I}(f; g)}$ is the **average** number of queries required to solve $f(x) = y_k$ when given access to $g(x)$. Since $\mathrm{I}(f; g) = \mathrm{E}(\mathrm{I}(f = y_k; g))$, these two kinds of estimation of complexity is related by Markov's inequality: *if $\xi$ is a nonnegative random variable, then* $\Pr(\xi \geq \varepsilon) \leq \dfrac{\mathrm{E}(\xi)}{\varepsilon}$ *for any* $\varepsilon > 0$. As we shall see, this inherent connection places an important restriction on both lower and upper bound on the time complexity of solving $f(x) = y_k$.

With this purpose, consider the following identity

$$\frac{\mathrm{I}(f = y_k)}{\mathrm{I}(f; g)} = \frac{\mathrm{I}(f = y_k)}{\mathrm{H}(f)} \times \frac{\mathrm{H}(f)}{\mathrm{I}(f; g)}. \tag{43}$$

This identity is natural in that $\dfrac{\mathrm{I}(f = y_k)}{\mathrm{H}(f)}$ is precisely the ratio of the amount of self-information about the event $\{f = y_k\}$ to the average amount of information contained in $f$. Also note that $\dfrac{\mathrm{H}(f)}{\mathrm{I}(f; g)}$ amounts to the **expected** number of queries required to invert $f(x)$ in case $f \leq_{\mathrm{T}} g$. Since $\mathrm{I}(f; g) \leq \min(\mathrm{H}(f), \mathrm{H}(g))$, we always have $\dfrac{\mathrm{H}(f)}{\mathrm{I}(f; g)} \geq 1$.

In fact, $\dfrac{\mathrm{I}(f = y_k)}{\mathrm{H}(f)}$ gives rise to a lower bound on the query complexity of solving $f(x) = y_k$ *among all possible Turing reductions*. To see this, we shall consider the efficiency of an *ideal* Turing reduction, in a way similar to Carnot's work on measuring the efficiency of an ideal heat engine in thermodynamics (see Feynman *et al.* [2013]). It is well known that the most efficient heat engine is an idealized engine in which all the processes are reversible. For any reversible engine that works between temperatures $T_1$ and $T_2$, the heat $Q_1$ absorbed at $T_1$ and the heat $Q_2$ delivered at $T_2$ must be related by thermodynamics entropy, that is,

$$\frac{Q_1}{T_1} = \frac{Q_2}{T_2} = entropy.$$  (44)

*No heat engine working between temperatures $T_1$ and $T_2$ can do more work than a reversible engine*. This principle reflects that the behavior of the nature must be limited in a certain way.

The same argument applies well to Turing reductions. To see this, consider a reversible Turing reduction $f = y_k \leq_T g = z_i$, that is, we assume $g = z_i$ is also reducible to $f = y_k$. In this case we say the search problem $f = y_k$ is *Turing equivalent* to $g = z_i$, denoted as $f = y_k \equiv_T g = z_i$. As a result, the **least** number of queries needed by solving $g(x) = z_i$ is equal to $time(g = z_i) = \dfrac{\mathrm{I}(g = z_i)}{\mathrm{I}(g = z_i; f = y_k)}$. By the symmetry of the pointwise mutual information, we obtain the following identity

$$\frac{\mathrm{I}(f = y_k)}{time(f = y_k)} = \mathrm{I}(f = y_k; g = z_i) = \mathrm{I}(g = z_i; f = y_k) = \frac{\mathrm{I}(g = z_i)}{time(g = z_i)}.$$  (45)

Further, if $f$ is *Turing equivalent* to $g$, then the **average** number of queries needed to solve $f = y_k$ and $g = z_i$ must be related by

$$\frac{\mathrm{I}(f = y_k)}{time(f = y_k)} = \frac{\mathrm{I}(g = z_i)}{time(g = z_i)} = \mathrm{I}(f; g).$$  (46)

This relation is optimal in that $f \equiv_T g$ means that the average mutual information $\mathrm{I}(f; g)$ is **maximal** among all possible Turing reductions from $f(x)$ to other functions, because $f(x)$ and $g(x)$ contain all useful information about each other.

In the ideal case of $\mathrm{I}(f; g) = \mathrm{H}(f)$, then we obtain a perfect analogy of an ideal heat engine

$$\frac{\mathrm{I}(f = y_k)}{time(f = y_k)} = \frac{\mathrm{I}(g = z_i)}{time(g = z_i)} = \mathrm{H}(f).$$  (47)

In this ideal case, $time(f = y_k) = \dfrac{\mathrm{I}(f = y_k)}{\mathrm{H}(f)}$ is **minimal** among all possible Turing reductions from $f(x) = y_k$ to other functions. *No algorithm can solve $f(x) = y_k$ with query complexity strictly less than* $\dfrac{\mathrm{I}(f = y_k)}{\mathrm{H}(f)}$. Once again, this limitation is the property of the nature, not a property of a particular problem.

In conclusion, working always with reversible Turing reductions, the query complexity of solving $f(x) = y_k$ *does not depend on the design of the Turing machine*. This is a perfect analogy of Carnot's brilliant conclusion: that the efficiency of a reversible heat engine *does not depend on the design of the heat engine* (see Feynman *et al.* [2013]).

Developing this spirit further, if $\dfrac{\mathrm{I}(f = y_k)}{\mathrm{H}(f)}$ is *exponential* in the size of input, then the search problem $f(x) = y_k$ can not be reduced to any other function within polynomial time. According to the Church–Turing Thesis, in this case $f(x) = y_k$ is less likely to have polynomial-time algorithm on any ordinary Turing machine. This important result can be accurately stated and rigorously proved on the basis of Markov's inequality:

**THEOREM 4.1.** *Let $f : X \to \Re$ be a simple function with values $y_1, y_2, \cdots, y_n$ for sufficiently large $n$. Given $k$, if $\dfrac{\mathrm{I}(f = y_k)}{\mathrm{H}(f)}$ is exponential in the size of input, then the probability of $f(x) = y_k$ having polynomial-time algorithm is a negative exponential in the size of input.*

**PROOF.** According to the Church–Turing Thesis, it suffices to consider any Turing reduction from the search problem $f(x) = y_k$ to other function $g(x)$. Note that the search problem $f(x) = y_k$ has polynomial-time algorithm implies that the query complexity $time(f = y_k)$ of solving $f(x) = y_k$ must be polynomial in the size of input.

Firstly, since $\mathrm{I}(f; g) \le \min(\mathrm{H}(f), \mathrm{H}(g))$, we have $\dfrac{\mathrm{I}(f = y_k)}{\mathrm{I}(f; g)} \ge \dfrac{\mathrm{I}(f = y_k)}{\mathrm{H}(f)}$. According to the assumption, $\dfrac{\mathrm{I}(f = y_k)}{\mathrm{I}(f; g)}$ is also exponential in the size of input.

Secondly, the least number of queries of solving $f(x) = y_k$ always satisfies

$$time(f = y_k) = \frac{\mathrm{I}(f = y_k)}{\mathrm{I}(f = y_k; g)} = \frac{\mathrm{I}(f = y_k)}{\mathrm{I}(f; g)} \times \frac{\mathrm{I}(f; g)}{\mathrm{I}(f = y_k; g)}. \tag{48}$$

Now that $time(f = y_k)$ is polynomial and $\dfrac{I(f = y_k)}{I(f;g)}$ is exponential in the size of input, it follows that $\dfrac{I(f;g)}{I(f = y_k;g)} \leq \dfrac{1}{\varepsilon}$ for some $\varepsilon > 0$ exponential in the size of input.

Finally, since $I(f;g) = E(I(f = y_k;g))$, Markov's inequality gives rise to our desired result: $\Pr(\dfrac{I(f;g)}{I(f = y_k;g)} \leq \dfrac{1}{\varepsilon}) = \Pr(I(f = y_k;g) \geq \varepsilon I(f;g)) \leq \dfrac{1}{\varepsilon} .\square$

**Note**: The assumption of large $n$ means that this result does not apply to decision problems. But this difficulty is not essential, since we can consider search problems instead and work within the class of function problems (see Rich [2007], section 28.10).

In fact, theorem 4.1 enables us to reduce the proof of $P \neq NP$ to the computation of the entropy of certain simple functions with finite values. To see this, just note that $P \neq NP$ if and only if $FP \neq FNP$ (see Rich [2007], section 28.10). In view of this fact, consider the search version of the subset sum problem: given a set of non-zero integers $a_1, a_2, \cdots, a_n \in Z$, **find** a non-empty subset that adds up to $0$ (see section 3.3 for subset sum decision problem). In terms of the subset sum function (see section 3.3), the subset sum search problem amounts to solving $SUM(x) = 0$ and hence naturally falls into the class $FNP$. So, to prove $FP \neq FNP$ it suffices to show that $\dfrac{I(SUM = 0)}{H(SUM)}$ is *exponential* in $n$. In this way, we have reduced the proof of $P \neq NP$ to the computation of $H(SUM)$. As a result, $P \neq NP$ is provable, answering a long-standing open problem (see S. Aaronson [2003]).

However, due to lack of information about the distribution of the values of arbitrary subset sum function, it seems hard to calculate $H(SUM)$ in general. For further discussions see section 5.1.

To conclude this subsection, we point out that there are several cases wroth noting:

Firstly, if the function $f$ takes values with equal probability, then $I(f = y_k) = H(f)$ for all $k$ and hence $\dfrac{I(f = y_k)}{I(f;g)} = \dfrac{H(f)}{I(f;g)}$ for any Turing reduction from $f(x) = y_k$ to $g(x)$. As a result, the average-case complexity of inverting $f(x)$ is as hard as the worst case of solving $f(x) = y_k$. For example, the discrete logarithm (*Example 3.5*), the quadratic residuosity (*Example 3.6*), and the RSA inversion problem (*Example 3.7*) all fall into this category. This result agrees with that based on Random Self-reducibility (see M. Abadi *et al.* [1989]).

Secondly, if the probability $p_k = 0$, then the self-information $I_k(f) = \infty$. In such a case, each query reveals *no* information about the solution, since there is no satisfying input at all. In part, this degeneracy can be attributed to the asymmetry of the class $NP$ (see Kleinberg and Tardos [2005], section 8.9). But, this degeneracy does not necessarily imply that deciding whether $f(x) = y_k$ has solution will be hard when there is *no* satisfying input. Instead, as far as the random $K$-SAT problem is concerned, when there is *no* satisfying truth assignment randomly generated $K$-CNF formulas become relatively easy to solve by showing them to be unsatisfiable (for details see Kirkpatrick and Selman [1994]; Selman *et al.* [1996]; Monasson *et al.* [1999]; Achlioptas and Peres [2004]).[7]

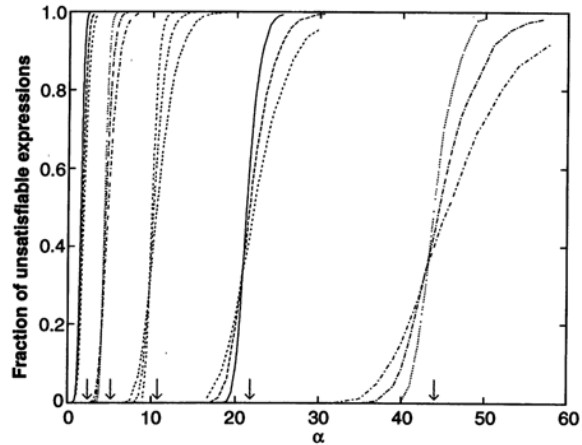

Fig. 7. Fraction of unsatisfiable expressions for K-SAT (from left to right, K=2 to 6).

*Source:* Kirkpatrick and Selman, Science, vol. 264, 1994.

---

[7] Kirkpatrick and Selman [1994] showed that random $K$-CNF formulas can exhibit sharp phase transition phenomenon as a function of the ratio of clauses to variables. For a random $K$-CNF formula $F$ with $n$ variables and $m$ clauses, its *density* $\dfrac{m}{n}$ determines what fraction of the randomly generated formulas is satisfiable. At low densities, the problem is *under-constrained* and almost all formulas are satisfiable, whereas at high densities the problem is *over-constrained* and almost all formulas are unsatisfiable. As the number of variables grows, the transition from almost always satisfiable to almost always unsatisfiable becomes discontinuous for $K \geq 3$ (see Achlioptas and Peres [2004]). It is an interesting and important fact that it is around the threshold that the computational cost is maximum, and the peak on computational cost sharpens up for higher values of $n$. But away from the threshold it becomes relatively easy to solve random formulas by showing them to be satisfiable at low densities and unsatisfiable at high densities respectively (see Selman *et al.* [1996]). In fact, it is at the "$50\%$ satisfiable" point where the computationally hardest instances are found. Intuitively, since random $K$-CNF formulas with few clauses are under-constrained and have many satisfying assignments, a truth assignment is likely to be found in the search. Formulas with very many clauses are over-constrained and usually unsatisfiable, so contradictions are found easily. As a result, the hardest area for satisfiability is near the point where, on average, only **one** satisfying assignment survives for half of random instances (see Kirkpatrick and Selman [1994]).

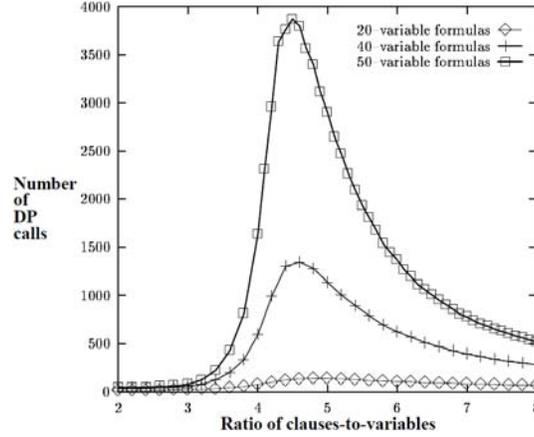

Fig. 8. Computational cost of satisfiability testing for 3-SAT.

*Source:* Selman *et al.*, Artificial Intelligence, vol. 81, 1996.

To justify the usefulness of our framework, we shall show how to get upper bound as well as lower bound on the time complexity of solving $f(x) = y_k$ within our framework.

### 4.2 Upper Bound

Given a *Turing reduction* from the search problem $f(x) = y_k$ to another function $g(x)$, in theory $\dfrac{\mathrm{I}(f = y_k)}{\mathrm{I}(f;g)}$ gives an **upper bound** on the query complexity of solving $f(x) = y_k$. In the idea case, if $\dfrac{\mathrm{I}(f = y_k)}{\mathrm{I}(f;g)}$ is polynomial in the size of input and the function $g(x)$ can be computed in polynomial time, then the search problem $f(x) = y_k$ also has polynomial-time algorithm.

Applying to decision problems, our information-based complexity theory is a natural setting in which to study the power of randomized or probabilistic algorithms of decision problems. To justify the usefulness of our framework, we shall consider the complexity of the class $BPP$, consisting of decision problems that have Bounded-error, Probabilistic, Polynomial time algorithms (see Rich [2007], section 30.2; Papadimitriou [1994]). There are two big unknowns concerning the class $BPP$. One unknown is the relationship between $BPP$ and $NP$. The other unknown is whether $P = BPP$, which is widely conjectured to be true.

*Example 4.1 (The class $BPP$).* Formally, a language $L$ over the alphabet $\{0,1\}$ belongs to the class $BPP$ if and only if there exists some probabilistic Turing machine that runs in polynomial time on all inputs and that decides $L$ with false probability $\varepsilon$ for arbitrarily small constant $0 < \varepsilon < \dfrac{1}{2}$.

Now assume that the characteristic function of $L$ is $\chi_L$ with the probability of acceptance $p_1$ and the probability of rejection $p_0$. Then the entropy of the given language $L$ equals $H(\chi_L) = -(p_0 \log p_0 + p_1 \log p_1)$.

Now $L \in BPP$ means that there are two kinds of error: to accept words that should be rejected $(0 \text{ to } 1)$, or to reject words that should be accepted $(1 \text{ to } 0)$.[8] This gives rise to a $\{0,1\}$-valued function $\chi_L^\varepsilon$ on $\{0,1\}^*$, with probabilities $\Pr(\chi_L^\varepsilon = 0) = p_0(1-\varepsilon) + p_1\varepsilon$ and $\Pr(\chi_L^\varepsilon = 1) = p_0\varepsilon + p_1(1-\varepsilon)$. So the entropy of the function $\chi_L^\varepsilon$ is

$$\mathrm{H}(\chi_L^\varepsilon) = [p_0(1-\varepsilon) + p_1\varepsilon]\log\frac{1}{p_0(1-\varepsilon)+p_1\varepsilon} + [p_0\varepsilon + p_1(1-\varepsilon)]\log\frac{1}{p_0\varepsilon + p_1(1-\varepsilon)} \quad (49)$$

Note that $\lim_{\varepsilon \to 0}\mathrm{H}(\chi_L^\varepsilon) = \mathrm{H}(\chi_L)$.

Next, we shall calculate the average mutual information via

$$\mathrm{I}(\chi_L; \chi_L^\varepsilon) = \mathrm{H}(\chi_L^\varepsilon) - \mathrm{H}(\chi_L^\varepsilon \mid \chi_L). \quad (50)$$

To do this, we need to compute the conditional entropy $\mathrm{H}(\chi_L^\varepsilon \mid \chi_L)$, which in turn needs the following conditional probabilities

$$\begin{aligned}
\Pr(\chi_L^\varepsilon = 0 \mid \chi_L = 0) = 1-\varepsilon, \quad & \Pr(\chi_L^\varepsilon = 1 \mid \chi_L = 0) = \varepsilon, \\
\Pr(\chi_L^\varepsilon = 0 \mid \chi_L = 1) = \varepsilon, \quad & \Pr(\chi_L^\varepsilon = 1 \mid \chi_L = 1) = 1-\varepsilon.
\end{aligned} \quad (51)$$

So the conditional entropy $\mathrm{H}(\chi_L^\varepsilon \mid \chi_L)$ is by definition equal to

$$\begin{aligned}
\mathrm{H}(\chi_L^\varepsilon \mid \chi_L) &= \Pr(\chi_L = 0)\mathrm{H}(\chi_L^\varepsilon \mid \chi_L = 0) + \Pr(\chi_L = 1)\mathrm{H}(\chi_L^\varepsilon \mid \chi_L = 1) \\
&= \Pr(\chi_L = 0)\Pr(\chi_L^\varepsilon = 0 \mid \chi_L = 0)\log\frac{1}{\Pr(\chi_L^\varepsilon = 0 \mid \chi_L = 0)} \\
&\quad + \Pr(\chi_L = 0)\Pr(\chi_L^\varepsilon = 1 \mid \chi_L = 0)\log\frac{1}{\Pr(\chi_L^\varepsilon = 1 \mid \chi_L = 0)} \\
&\quad + \Pr(\chi_L = 1)\Pr(\chi_L^\varepsilon = 0 \mid \chi_L = 1)\log\frac{1}{\Pr(\chi_L^\varepsilon = 0 \mid \chi_L = 1)} \\
&\quad + \Pr(\chi_L = 1)\Pr(\chi_L^\varepsilon = 1 \mid \chi_L = 1)\log\frac{1}{\Pr(\chi_L^\varepsilon = 1 \mid \chi_L = 1)} \\
&= p_0[(1-\varepsilon)\log\frac{1}{1-\varepsilon} + \varepsilon\log\frac{1}{\varepsilon}] + p_1[(1-\varepsilon)\log\frac{1}{1-\varepsilon} + \varepsilon\log\frac{1}{\varepsilon}] \\
&= (1-\varepsilon)\log\frac{1}{1-\varepsilon} + \varepsilon\log\frac{1}{\varepsilon} = -[(1-\varepsilon)\log(1-\varepsilon) + \varepsilon\log\varepsilon]
\end{aligned} \quad (52)$$

---

As a result, the average mutual information of $\chi_L$ and $\chi_L^\varepsilon$ equals

$$\begin{aligned}
\mathrm{I}(\chi_L;\chi_L^\varepsilon) &= \mathrm{H}(\chi_L^\varepsilon) - \mathrm{H}(\chi_L^\varepsilon \mid \chi_L) \\
&= [\,p_0(1-\varepsilon) + p_1\varepsilon\,]\log\frac{1}{p_0(1-\varepsilon) + p_1\varepsilon} \\
&\quad + [\,p_0\varepsilon + p_1(1-\varepsilon)\,]\log\frac{1}{p_0\varepsilon + p_1(1-\varepsilon)} \\
&\quad - [(1-\varepsilon)\log\frac{1}{1-\varepsilon} + \varepsilon\log\frac{1}{\varepsilon}]
\end{aligned} \tag{53}$$

Note that the mutual information $\mathrm{I}(\chi_L;\chi_L^\varepsilon)$ is completely determined by the acceptance probability $p_1$, which is an attribute the language, and the false probability $\varepsilon$, which is an attribute of the probabilistic Turing machine.

By the **CONTINUITY** properties of the entropy measure, we have $\lim_{\varepsilon\to 0}\mathrm{I}(\chi_L;\chi_L^\varepsilon) = H(\chi_L)$. Consequently, given language $L \in BPP$, $\dfrac{\mathrm{H}(\chi_L)}{\mathrm{I}(\chi_L;\chi_L^\varepsilon)}$ approaches $1$ as $\varepsilon\to 0$.

On the other hand, the **expected** number of queries satisfy

$$\frac{\mathrm{I}(\chi_L = 1)}{\mathrm{I}(\chi_L;\chi_L^\varepsilon)} = \frac{\mathrm{I}(\chi_L = 1)}{\mathrm{I}(\chi_L^\varepsilon = 1)} \times \frac{\mathrm{I}(\chi_L^\varepsilon = 1)}{\mathrm{H}(\chi_L^\varepsilon)} \times \frac{\mathrm{H}(\chi_L^\varepsilon)}{\mathrm{H}(\chi_L)} \times \frac{\mathrm{H}(\chi_L)}{\mathrm{I}(\chi_L;\chi_L^\varepsilon)}. \tag{54}$$

Consequently, we have $\dfrac{\mathrm{I}(\chi_L = 1)}{\mathrm{I}(\chi_L;\chi_L^\varepsilon)} \approx \dfrac{\mathrm{I}(\chi_L^\varepsilon = 1)}{\mathrm{H}(\chi_L^\varepsilon)}$ as $\varepsilon\to 0$.

Since $L \in BPP$, the decision problem $\chi_L^\varepsilon = 1$ has polynomial time algorithm according to the definition of $BPP$, and hence its query complexity $\dfrac{\mathrm{I}(\chi_L^\varepsilon = 1)}{\mathrm{H}(\chi_L^\varepsilon)}$ is polynomial in the size of input. As such, any language $L$ in the class $BPP$ can be decided by a deterministic Turing machine within polynomial time. This is a strong evidence to support $P = RP = BPP$.

**Note**: This result does not necessarily imply that the repetition of a probabilistic, polynomial time algorithm will itself become into an efficient algorithm. A typical example is the famous Primality Test (see Papadimitriou [1994]). It is well known that the Miller–Rabin algorithm runs in polynomial time and has an error rate that is less than an inverse exponential (see Rabin [1980]). Its original version, the Miller test, is fully deterministic and runs in polynomial time over all inputs, but its correctness relies on the Extended Riemann hypothesis (see Miller [1976]). In contrast, Agrawal, Kayal and Saxena [2004] have created an unconditional deterministic polynomial-time algorithm to determine whether a natural number is prime and shown that Primality test turns out to be in the class $P$. $\square$

### 4.3   Lower Bound

To obtain a lower bound on the time complexity of solving $f(x) = y_k$ we have to consider the *tradeoff* between the number of queries and the complexity of each query. Given a *Turing reduction* from the search problem $f(x) = y_k$ to another function $g(x)$, if we can estimate the average mutual information $\mathrm{I}(f;g)$, then we can estimate the **average** number of queries needed by solving $f(x) = y_k$. Extending this further, by knowing upper or lower bound on the average mutual information $\mathrm{I}(f;g)$, we can get lower or upper bound on time complexity of solving $f(x) = y_k$ respectively (for similar ideas see Shah and Sharma [2000]).

One way to obtain lower bound on the time complexity is based on the properties of the entropy. Since $\mathrm{I}(f;g) \leq \mathrm{H}(f)$, the **average** number of queries needed by solving $f(x) = y_k$ will satisfy

$$\frac{\mathrm{I}(f = y_k)}{\mathrm{I}(f;g)} \geq \frac{\mathrm{I}(f = y_k)}{\mathrm{H}(f)}. \tag{55}$$

In the worst case, if $\dfrac{\mathrm{I}(f = y_k)}{\mathrm{H}(f)}$ is exponential in the size of input, then the search problem $f(x) = y_k$ is less likely to have polynomial-time algorithm.

Applying to the famous Boolean Satisfiability problem, our result supports that $P \neq NP$. Recall that Boolean Satisfiability problem is the first problem that was proven to be $NP$-complete, independently by Cook [1971] and Levin [1973].

*Example 4.2 (Boolean Satisfiability Problem, SAT).* In terms of Boolean functions it is possible to express the SAT problem as follows: Given a Boolean function $BOOL : \{0,1\}^n \to \{0,1\}$, decide whether there exists $x \in \{0,1\}^n$ such that $BOOL(x) = 1$.

Now given a Boolean function $BOOL : \{0,1\}^n \to \{0,1\}$, the self-information associated with $\{BOOL = 1\}$ is $\mathrm{I}_1(BOOL) = -\log p_1$ and the entropy of $BOOL$ is $\mathrm{H}(BOOL) = -(p_0 \log p_0 + p_1 \log p_1)$, so the **average** number of trials of deciding whether $BOOL = 1$ is satisfiable is no less than

$$\frac{\mathrm{I}(BOOL = 1)}{\mathrm{H}(BOOL)} = \frac{-\log p_1}{-p_1 \log p_1 - (1 - p_1)\log(1 - p_1)}. \tag{56}$$

Suppose that $p_1 = \dfrac{k}{2^n}$ for some positive number $k$ far less than $2^n$, then the **average** number of trials for determining whether $BOOL = 1$ is at least

$$\frac{\mathrm{I}(BOOL=1)}{\mathrm{H}(BOOL)} = \frac{-\log \frac{k}{2^n}}{-\frac{k}{2^n}\log \frac{k}{2^n} - (1-\frac{k}{2^n})\log(1-\frac{k}{2^n})}. \tag{57}$$

Since $\log(1-\frac{k}{2^n}) \approx -\frac{k}{2^n}$ , we have

$$\frac{\mathrm{I}(BOOL=1)}{\mathrm{H}(BOOL)} \approx \frac{n-\log k}{\frac{k}{2^n}(n-\log k) + \frac{k}{2^n}(1-\frac{k}{2^n})} = \frac{n-\log k}{n-\log k + (1-\frac{k}{2^n})} \times \frac{2^n}{k}. \tag{58}$$

It is easy to see that $\frac{\mathrm{I}(BOOL=1)}{\mathrm{H}(BOOL)}$ is approximately equal to $\frac{2^n}{k} = \frac{1}{p_1}$ as $k$ is

far less than $2^n$. It is worth noting that $\frac{1}{p_1}$ is exactly the **expected** number of

Bernoulli trials until the *first* success. Consequently, the maximum number of trials will be $2^n$, and this maximum holds only when there is only **one** satisfying assignment survives ($p_1 = \frac{1}{2^n}$).

This result is consistent with existing results for random $K$-SAT problem. For $K \geq 3$, random $K$-CNF formulas exhibits sharp phase transitions near the point where only **one** satisfying assignment survives for half of random instances and the computational cost grow exponentially with problem size near the transition threshold (for details see Kirkpatrick and Selman [1994]; Selman *et al.* [1996]; Monasson *et al.* [1999]; Achlioptas and Peres [2004]). □

Another way to obtain a lower bound on the time complexity is to use brute-force search as a benchmark for efficiency (see Kleinberg and Tardos [2005]). To do this, just note that to solve $f(x) = y_k$ only *one* satisfying $x$ is needed to be found, rather than finding out *all*. So any algorithm for solving $f(x) = y_k$ will stop when the *first* solution was found. It is well known that the **expected** number of Bernoulli trials until the first success can be modeled by the geometric distribution. Formally, if the probability of success on each trial is $p_k$, then the expected value of Bernoulli trials until the *first* success is $\frac{1}{p_k}$.

Therefore, using brute-force search as a benchmark for efficiency, we can get a natural upper bound on the **average** number of queries as follows

$$\frac{\mathrm{I}(f=y_k)}{\mathrm{I}(f;g)} \leq \frac{1}{p_k}. \tag{59}$$

Or equivalently,

$$\mathrm{I}(f;g) \geq p_k \mathrm{I}(f = y_k).\tag{60}$$

This inequality provides an absolute limit on the average mutual information for any reduction-based algorithm that is more efficient than brute-force search. Intuitively, it can be interpreted as that the amount of information that $g(x)$ contains about $f(x) = y_k$ can not be less than the average amount of average self-information associated with the event $\{f = y_k\}$.

Otherwise, if $\mathrm{I}(f;g) < p_k \mathrm{I}(f = y_k)$, then brute-force search will be more efficient in solving $f(x) = y_k$ than any reduction-based algorithm. In such a case, the **average** number of queries will satisfy $\dfrac{\mathrm{I}(f = y_k)}{\mathrm{I}(f;g)} > \dfrac{1}{p_k}$. In the worst case, if $\dfrac{1}{p_k}$ is exponential in the size of input, then the search problem $f(x) = y_k$ is less likely to have polynomial-time algorithm.

Applying to decision problems, we shall consider the complexity of the class $PP$, the class of decision problems solvable by Polynomial-time Probabilistic algorithms that decide whether to accept "by majority" (see Papadimitriou [1994]). Formally, a language $L$ over the alphabet $\{0,1\}$ belongs to the class $PP$ if and only if there exists some probabilistic Turing machine $M$ that runs in polynomial time on all inputs and that decides $L$ with false probability of less than $\dfrac{1}{2}$. That is, if $w \notin L$ then $M$ rejects $w$ with probability $\dfrac{1}{2} + \varepsilon$, and if $w \in L$ then $M$ accepts $w$ with probability $\dfrac{1}{2} + \varepsilon$, where $0 < \varepsilon < \dfrac{1}{2}$ may depend on the size of inputs.

A classical problem in the class $PP$ is the following MAJORITY SAT: Given a Boolean expression of $n$ variables, or equivalently, a Boolean function from $\{0,1\}^n$ to $\{0,1\}$, decide whether the majority of the $2^n$ truth assignment to its variables (that is, at least $2^{n-1} + 1$ of them) satisfies it? In this case $\varepsilon = \dfrac{1}{2^n}$, with just two more accepting computations than rejecting computations.

Firstly, we shall show that our information-based estimation of repetitions of polynomial-time probabilistic algorithm is superior to that based on the Chernoff bound (see Papadimitriou [1994]).

*Example 4.3 (The class $PP$).* Now assume the characteristic function of $L$ to be $\chi_L$ with the probability of acceptance $p_1$ and the probability of rejection $p_0$. Then the entropy of the given language $L$ equals $\mathrm{H}(\chi_L) = -(p_0 \log p_0 + p_1 \log p_1)$.

On the other hand, $L \in PP$ gives rise to a function $\chi_L^\varepsilon$ from $\{0,1\}^*$ to the set $\{0,1\}$ with probabilities

$$
\begin{aligned}
\Pr(\chi_L^\varepsilon = 0) = p_0(\frac{1}{2} + \varepsilon) + p_1(\frac{1}{2} - \varepsilon) = \frac{1}{2} + \varepsilon(p_0 - p_1) \\
\Pr(\chi_L^\varepsilon = 1) = p_0(\frac{1}{2} - \varepsilon) + p_1(\frac{1}{2} + \varepsilon) = \frac{1}{2} + \varepsilon(p_1 - p_0)
\end{aligned}
. \tag{61}
$$

So the entropy of the function $\chi_L^\varepsilon$ is

$$
H(\chi_L^\varepsilon) = [\frac{1}{2} + \varepsilon(p_0 - p_1)]\log \frac{1}{\frac{1}{2} + \varepsilon(p_0 - p_1)} + [\frac{1}{2} + \varepsilon(p_1 - p_0)]\log \frac{1}{\frac{1}{2} + \varepsilon(p_1 - p_0)} \tag{62}
$$

The language $L \in PP$ yields the following conditional probabilities

$$
\begin{aligned}
\Pr(\chi_L^\varepsilon = 0 \mid \chi_L = 0) = \frac{1}{2} + \varepsilon, \quad \Pr(\chi_L^\varepsilon = 1 \mid \chi_L = 0) = \frac{1}{2} - \varepsilon, \\
\Pr(\chi_L^\varepsilon = 0 \mid \chi_L = 1) = \frac{1}{2} - \varepsilon, \quad \Pr(\chi_L^\varepsilon = 1 \mid \chi_L = 1) = \frac{1}{2} + \varepsilon.
\end{aligned}
\tag{63}
$$

So the conditional entropy $H(\chi_L^\varepsilon \mid \chi_L)$ is by definition equal to

$$\mathrm{H}(\chi_L^\varepsilon \mid \chi_L) = \Pr(\chi_L = 0)\mathrm{H}(\chi_L^\varepsilon \mid \chi_L = 0) + \Pr(\chi_L = 1)\mathrm{H}(\chi_L^\varepsilon \mid \chi_L = 1)$$

$$= \Pr(\chi_L = 0)\Pr(\chi_L^\varepsilon = 0 \mid \chi_L = 0)\log \frac{1}{\Pr(\chi_L^\varepsilon = 0 \mid \chi_L = 0)}$$

$$+ \Pr(\chi_L = 0)\Pr(\chi_L^\varepsilon = 1 \mid \chi_L = 0)\log \frac{1}{\Pr(\chi_L^\varepsilon = 1 \mid \chi_L = 0)}$$

$$+ \Pr(\chi_L = 1)\Pr(\chi_L^\varepsilon = 0 \mid \chi_L = 1)\log \frac{1}{\Pr(\chi_L^\varepsilon = 0 \mid \chi_L = 1)}$$

$$+ \Pr(\chi_L = 1)\Pr(\chi_L^\varepsilon = 1 \mid \chi_L = 1)\log \frac{1}{\Pr(\chi_L^\varepsilon = 1 \mid \chi_L = 1)}$$

$$= p_0[(\frac{1}{2}+\varepsilon)\log \frac{1}{\frac{1}{2}+\varepsilon} + (\frac{1}{2}-\varepsilon)\log \frac{1}{\frac{1}{2}-\varepsilon}]$$

$$+ p_1[(\frac{1}{2}-\varepsilon)\log \frac{1}{\frac{1}{2}-\varepsilon} + (\frac{1}{2}+\varepsilon)\log \frac{1}{\frac{1}{2}+\varepsilon}]$$

$$= (\frac{1}{2}-\varepsilon)\log \frac{1}{\frac{1}{2}-\varepsilon} + (\frac{1}{2}+\varepsilon)\log \frac{1}{\frac{1}{2}+\varepsilon} \qquad\qquad (64)$$

$$= -[(\frac{1}{2}-\varepsilon)\log \frac{1}{2}(1-2\varepsilon) + (\frac{1}{2}+\varepsilon)\log \frac{1}{2}(1+2\varepsilon)]$$

Now that $\lim_{\varepsilon \to 0} \mathrm{H}(\chi_L^\varepsilon) = \lim_{\varepsilon \to 0} \mathrm{H}(\chi_L^\varepsilon \mid \chi_L) = 1$, the mutual information of $\chi_L$ and $\chi_L^\varepsilon$ satisfies

$$\lim_{\varepsilon \to 0} \mathrm{I}(\chi_L; \chi_L^\varepsilon) = \lim_{\varepsilon \to 0} \mathrm{H}(\chi_L^\varepsilon) - \lim_{\varepsilon \to 0} \mathrm{H}(\chi_L^\varepsilon \mid \chi_L) = 0. \qquad (65)$$

In such a case, each query of $\chi_L^\varepsilon$ reveals nearly *no* information about the language. This means that in the worst case an *exponential* number of repetitions of the probabilistic algorithm may be required in order to determine the correct answer with reasonable confidence (see Papadimitriou [1994]).

To see this, note that $\log(1 + x) \approx x$ for $|x| < 1$. Consequently, it is routine to check that

$$\mathrm{H}(\chi_L^{\varepsilon} \mid \chi_L) = -[(\frac{1}{2} - \varepsilon)\log\frac{1}{2}(1 - 2\varepsilon) + (\frac{1}{2} + \varepsilon)\log\frac{1}{2}(1 + 2\varepsilon)]$$

$$= 1 - [(\frac{1}{2} - \varepsilon)\log(1 - 2\varepsilon) + (\frac{1}{2} + \varepsilon)\log(1 + 2\varepsilon)]$$

$$\approx 1 - [(\frac{1}{2} - \varepsilon) \times (-2\varepsilon) + (\frac{1}{2} + \varepsilon) \times (2\varepsilon)]$$ 　(66)

$$= 1 - 4\varepsilon^2$$

and

$$\mathrm{H}(\chi_L^{\varepsilon}) = [\frac{1}{2} + \varepsilon(p_0 - p_1)]\log\frac{1}{\frac{1}{2} + \varepsilon(p_0 - p_1)} + [\frac{1}{2} + \varepsilon(p_1 - p_0)]\log\frac{1}{\frac{1}{2} + \varepsilon(p_1 - p_0)}$$

$$= -[\frac{1}{2} + \varepsilon(p_0 - p_1)]\log\frac{1}{2}[1 + 2\varepsilon(p_0 - p_1)] - [\frac{1}{2} + \varepsilon(p_1 - p_0)]\log\frac{1}{2}[1 + 2\varepsilon(p_1 - p_0)]$$

$$= 1 - [\frac{1}{2} + \varepsilon(p_0 - p_1)]\log[1 + 2\varepsilon(p_0 - p_1)] - [\frac{1}{2} + \varepsilon(p_1 - p_0)]\log[1 + 2\varepsilon(p_1 - p_0)]$$ 　.(67)

$$\approx 1 - [\frac{1}{2} + \varepsilon(p_0 - p_1)] \times 2\varepsilon(p_0 - p_1) - [\frac{1}{2} + \varepsilon(p_1 - p_0)] \times 2\varepsilon(p_1 - p_0)$$

$$= 1 - 4\varepsilon^2(p_0 - p_1)^2$$

Consequently, we obtain an approximation of the average mutual information

$$\mathrm{I}(\chi_L; \chi_L^{\varepsilon}) = \mathrm{H}(\chi_L^{\varepsilon}) - \mathrm{H}(\chi_L^{\varepsilon} \mid \chi_L) \approx 4\varepsilon^2[1 - (p_0 - p_1)^2].$$ 　(68)

As a result, the **expected** number of repetitions of the probabilistic algorithm is approximately equal to

$$\frac{\mathrm{H}(\chi_L)}{\mathrm{I}(\chi_L; \chi_L^{\varepsilon})} \approx \frac{1}{\varepsilon^2} \times \frac{\mathrm{H}(\chi_L)}{4[1 - (p_0 - p_1)^2]} = \frac{1}{\varepsilon^2} \times \frac{\mathrm{H}(\chi_L)}{16 p_0 p_1}.$$ 　(69)

It is easy to see that this exponential bound on the repetitions of the algorithm is superior to that based on the Chernoff bound (see Papadimitriou [1994]). □

Secondly, we shall show that our estimation of upper bound on the **average** number of queries provides strong evidence that $P \neq PP$.

*Example 4.4 (The class PP, continued).* Now that the probability of acceptance is $p_1$, the upper bound on the **average** number of queries yields the following inequality

$$\mathrm{I}(\chi_L; \chi_L^{\varepsilon}) \geq p_1 \mathrm{I}(\chi_L = 1) = -p_1 \log p_1.$$ 　(70)

This inequality provides an absolute limit on the error factor $\varepsilon$. Intuitively, to get an efficient algorithm by repeating the given probabilistic algorithm, there must be a lower bound on the error factor. In theory, this lower bound of the error factor ($\varepsilon$) can be solved from inequality (70) as a function of the probability of acceptance ($p_1$). However, due to its **nonlinearity**, analytic solution is hard to be found.

To overcome this difficulty, we shall try approximation. To do this, recall that we have got an approximation of the average mutual information

$$\mathrm{I}(\chi_L; \chi_L^\varepsilon) \approx 4\varepsilon^2[1 - (p_0 - p_1)^2] = 16\varepsilon^2 p_0 p_1. \tag{71}$$

On the other hand, we have

$$\mathrm{I}(\chi_L = 1) = -\log p_1 = -\log(1 - p_0) \approx p_0. \tag{72}$$

Taking together, we get an approximation of the average mutual information, that is,

$$\mathrm{I}(\chi_L; \chi_L^\varepsilon) \geq p_1 \mathrm{I}(\chi_L = 1) \Rightarrow 4\varepsilon^2[1 - (p_0 - p_1)^2] \geq p_0 p_1. \tag{73}$$

Or equivalently,

$$\mathrm{I}(\chi_L; \chi_L^\varepsilon) \geq p_1 \mathrm{I}(\chi_L = 1) \Rightarrow \varepsilon \geq \frac{1}{4}. \tag{74}$$

But, the existence of a lower bound on the error factor contradicts with the definition of the class $PP$. This contradiction supports that $P \neq PP$ as desired.

Otherwise, if the error factor $\varepsilon < \frac{1}{4}$, then brute-force search will be more efficient in deciding the language $L$ any reduction-based algorithm. In such a case, the class $PP$ is even less likely to coincide with $P$.

**Note**: This lower bound on error factor is highly inaccurate and is just used to illustrate the fundamental idea. The accurate bound on error factor ($\varepsilon$) can only be obtained by solving inequality (70) as a function of the probability of acceptance ($p_1$). But the idea is essentially the same.□

## 5.  OPEN PROBLEMS

In this section, we shall describe several open problems worthy of further studies.

### 5.1   The computation of Entropy

Being an attribute of measurable functions, the entropy $\mathrm{H}(f)$ conveys important information about $f : X \to \Re$. On the other hand, the computation of the entropy may also need some important information about the function in question. For example, it is shown that calculating the entropy of a Boolean function is at least as difficult as the Boolean satisfiability problem (example 3.3).

Further, applying an argument similar to Carnot's work on measuring the efficiency of an ideal heat engine in thermodynamics, it is shown that $\dfrac{\mathrm{I}(f = y_k)}{\mathrm{H}(f)}$ is **minimal** among all possible Turing reductions from $f(x) = y_k$ to other functions. *No algorithm can solve $f(x) = y_k$ with query complexity strictly less than* $\dfrac{\mathrm{I}(f = y_k)}{\mathrm{H}(f)}$. In the extreme case, if $\dfrac{\mathrm{I}(f = y_k)}{\mathrm{H}(f)}$ is *exponential* in the size of input, then the search problem $f(x) = y_k$ can not be reduced to any other function within polynomial time. According to the Church–Turing Thesis, in this case $f(x) = y_k$ is less likely to have polynomial-time algorithm on any ordinary Turing machine. According to Markov's inequality, *if $\dfrac{\mathrm{I}(f = y_k)}{\mathrm{H}(f)}$ is exponential in the size of input, then the probability of $f(x) = y_k$ having polynomial-time algorithm is a negative exponential in the size of input* (Theorem 4.1).

Unfortunately, the assumption of large $n$ means that theorem 4.1 does not apply to decision problems. But this difficulty is not essential, since we can consider search problems instead and work within the class of function problems (see Rich [2007], section 28.10). In fact, theorem 4.1 enables us to reduce the proof of $FP \neq FNP$ to the computation of the entropy of certain simple functions with finite values, such as the subset sum function. As a consequence of this reduction, $P \neq NP$ is provable. This result answers a long-standing open problem concerning this problem (see S. Aaronson [2003]).

However, due to lack of information about the distribution of the values of arbitrary subset sum function, it seems hard to calculate entropy of arbitrary subset sum function.

This naturally leads to the question of how to compute or approximately estimate the entropy of a given simple function with finite values, like Euler's totient function (example 3.4) or the integer multiplication function (example 3.8).

**5.2   Turing reduction between Functions**

In this paper, we have considered the case in which a search problem $f(x) = y_k$ is Turing reducible to another search problem $g(x) = z_i$, denoted by $f = y_k \leq_{\mathrm{T}} g = z_i$. Generally, this reduction is enabled by the information hidden in the function $f : X \to \Re$ and the information conveyed by the given value $y_k$.

It is shown that the **average** number of queries is equal to $\dfrac{\mathrm{I}(f = y_k)}{\mathrm{I}(f; g)}$.

However, $f = y_k \leq_{\mathrm{T}} g = z_i$ does not necessarily imply that the function $f(x)$ is Turing reducible to $g(x)$, denoted by $f \leq_{\mathrm{T}} g$ (see Ding-Zhu Du, Ker-I Ko [2013]). If $f \leq_{\mathrm{T}} g$, then for any $x \in X$, $f(x)$ can be computed by a Turing machine when given access to the oracle $g$. If both functions are Lebesgue measurable, then the **expected** number of queries exactly equals $\dfrac{\mathrm{H}(f)}{\mathrm{I}(f;g)}$.

Further, the fact that $f(x)$ can be computed by making queries to the oracle $g$ means that there must be an inherent connection between $f(x)$ and $g(x)$. In view of this, a natural question arises: How to characterize Turing reductions by means of the property of functions in question? A positive answer to this question may help us open the "black box" of oracle Turing machines (see Kleinberg and Tardos [2005]).

To illustrate, consider the set of Lebesgue measurable functions decided by a deterministic Turing machine with oracle access to given function $g : X \to \Re$, denoted by $L^g$. How to characterize the class $L^g$ by means of properties of the function $g$?

A question of interest is whether $L^g$ is invariant under *homotopy*? That is, if the given function $g : X \to \Re$ is *homotopic to* another function $f : X \to \Re$, how to characterize the relationship between $L^g$ and $L^f$? Recall that in this case there is a family of functions $h_t : X \to \Re$, one for each point $t \in [0,1]$, with $h_0 = g$ and $h_1 = f$, and the property that the map $(x,t) \mapsto h_t(x)$ is continuous from $X \times [0,1]$ to $\Re$ (see Armstrong [1983]).

### 5.3 Information Measure of Equations

Since the information of solutions of given equation must be contained in the equation itself, the information measure of equations may also be related to other important problems.

*Example 5.1 (Differential equations).* Consider a differential equation of the first order

$$\frac{dy}{dx} = f(x). \tag{75}$$

If $f(x)$ is Lebesgue measurable, then this differential equation determines one antiderivative $y = F(x)$, up to a constant of integration (see Patrick Billingsley [1995]). In this case, the entropy of the derivative $f(x)$ is well-defined and can be calculated by partitioning the range of $f(x)$ in the same way as computing its Lebesgue integral (see section 3.1). In turn, the entropy of the derivative $f(x)$ contains important information about the antiderivative $F(x)$.

On the other hand, Liouville's theorem states that the antiderivatives of certain elementary functions cannot themselves be expressed as elementary functions. A typical example is when $f(x) = e^{-x^2}$. Within our framework, it seems that this phenomenon may be attributed to the amount of information contained in the antiderivative in that being an elementary function places an important restriction on the amount of informatin contained in it.□

*Example 5.2 (Algebraic equations).* A polynomial equation of degree $n$ is an equation of the form

$$a_n x^n + a_{n-1} x^{n-1} + \cdots + a_1 x + a_0 = 0 \,. \tag{76}$$

The fundamental theorem of algebra states that every polynomial equation of degree $n$ has exactly $n$ roots, counted with multiplicity. On the other hand, the celebrated Abel–Ruffini theorem states that, for $n \geq 5$ there is no algebraic solution to polynomial equations with arbitrary coefficients. Galois independently proved the theorem by establishing a connection between field theory and group theory (see Jacobson [1974]).

Within our framework, it seems that the Abel–Ruffini–Galois impossibility theorem may have connection with the amount of information contained in a given equation. Intuitively, a polynomial function with higher degree usually behaves in a more complex way, which in turn means that their zero points distribute in a more random way. As a result, polynomial equations with higher degree usually contain more information.

To illustrate, consider a quadratic equation

$$ax^2 + bx + c = 0 \,. \tag{77}$$

If $a \neq 0$, then the quadratic formula gives rise to a vector function from $\Re^3$ to $\Re^2$

$$(a,b,c) \mapsto (\frac{-b - \sqrt{b^2 - 4ac}}{2a}, \frac{-b + \sqrt{b^2 - 4ac}}{2a}) \,. \tag{78}$$

Then the information contained in the quadratic equation can be defined to be the information contained in this vector function. □

*Example 5.3 (System of linear equations).* A general system of $m$ linear equations with $n$ unknowns can be written as a matrix equation of the form

$$A_{m \times n} x = b \,. \tag{79}$$

If the coefficient matrix is square ($m = n$) and has full rank, then the system has a unique solution given by

$$x = A^{-1} b \,. \tag{80}$$

In this case, we obtain a vector function $(A,b) \mapsto A^{-1} b$ from $\Re^{n^2+n}$ to $\Re^n$.

It seems that the amount of information contained in this function may be related to the time complexity of algorithms for solving $Ax = b$. Indeed, one of the biggest unsolved problems in numerical analysis is whether the complexity of solving a linear system $A_{n \times n} x = b$ can be reduced to $O(n^2)$ (see Trefethen [2012]). □

*Example 5.4 (Elliptic curves).* An elliptic curve can be written as a plane algebraic curve defined by a Weierstrass equation over a (finite) field

$$y^2 = x^3 + ax + b. \tag{81}$$

The use of elliptic curves in cryptography was suggested independently by Neal Koblitz [1987] and Miller [1985]. The amount of information contained in an elliptic curve may be important to Elliptic curve cryptography.□

## 6. CONCLUSIONS

The concept of Shannon entropy of random variables was generalized to measurable functions in general, and to simple functions with finite values in particular. The key point is to approximate a measurable function $f : X \to \Re$ by a simple function with finite values, which is obtained by partitioning the range of $f(x)$ in the same way as computing its Lebesgue integral. As a result, these values, coupled with the probability measure of their preimages, forms a random variable in itself. So, it is natural to approximate the entropy of the given function by the entropy of the resulting random variable.

It turns out that the information measure of measurable functions is related to the time complexity of search problems concerning the function in question. Formally, given a **Turing reduction** from a search problem $f(x) = y$ to another search problem $g(x) = z$, the amount of information about $f(x) = y$ provided by querying the solution(s) of $g(x) = z$ is exactly equal to the pointwise mutual information $I(f = y; g = z)$. As a result, the **least** number of queries is $time(f = y) = \dfrac{I(f = y)}{I(f = y; g = z)}$, where $I(f = y)$ is the amount of self-information about the event $\{f = y\}$.

Further, if the search problem $f(x) = y$ is Turing reducible to another function $g(x)$, then the **least** number of queries required to solve $f(x) = y$ will be $time(f = y) = \dfrac{I(f = y)}{I(f = y; g)}$, where $I(f = y; g)$ is the mutual information between the event $f(x) = y_k$ and the function $g(x)$. However, due to lack of information about the corresponding conditional probabilities, it seems difficult to analysis the **least** number of queries theoretically.

To avoid this logical difficulty, we shall consider the **average** number of queries instead. To this end, rewrite $time(f = y)$ as follows

$$time(f = y) = \frac{I(f = y)}{I(f = y; g)} = \frac{I(f = y)}{I(f; g)} \times \frac{I(f; g)}{I(f = y; g)}.$$

Since $\mathrm{I}(f;g)$ is the average amount of information between $f(x)$ and $g(x)$, $\dfrac{\mathrm{I}(f=y)}{\mathrm{I}(f;g)}$ is the **average** number of queries required to solve $f(x) = y$ when given access to $g(x)$. As we shall see, this inherent connection places an important restriction on both lower and upper bound on the time complexity of solving $f(x) = y$.

With this purpose, consider the following identity

$$\frac{\mathrm{I}(f=y)}{\mathrm{I}(f;g)} = \frac{\mathrm{I}(f=y)}{\mathrm{H}(f)} \times \frac{\mathrm{H}(f)}{\mathrm{I}(f;g)}.$$

This identity is natural in that $\dfrac{\mathrm{I}(f=y)}{\mathrm{H}(f)}$ is precisely the ratio of the amount of self-information about the event $\{f = y\}$ to the average amount of information contained in $f$. Also note that $\dfrac{\mathrm{H}(f)}{\mathrm{I}(f;g)}$ amounts to the **expected** number of queries required to invert $f(x)$ in case $f \leq_{\mathrm{T}} g$. Since $\mathrm{I}(f;g) \leq \min(\mathrm{H}(f), \mathrm{H}(g))$, we always have $\dfrac{\mathrm{H}(f)}{\mathrm{I}(f;g)} \geq 1$.

In fact, it turns out that $\dfrac{\mathrm{I}(f=y)}{\mathrm{H}(f)}$ is a lower bound on the query complexity of solving $f(x) = y$ among all possible Turing reductions. To see this, we shall consider the efficiency of an *ideal* Turing reduction, in a way similar to Carnot's work on measuring the efficiency of an ideal heat engine in thermodynamics (see Feynman *et al.* [2013]). It is well known that the most efficient heat engine is an idealized engine in which all the processes are reversible. For any reversible engine that works between temperatures $T_1$ and $T_2$, the heat $Q_1$ absorbed at $T_1$ and the heat $Q_2$ delivered at $T_2$ must be related by thermodynamics entropy, that is,

$$\frac{Q_1}{T_1} = \frac{Q_2}{T_2} = entropy.$$

*No heat engine working between temperatures $T_1$ and $T_2$ can do more work than a reversible engine.* This principle reflects that the behavior of the nature must be limited in a certain way.

The same argument applies well to reversible Turing reductions. To see this, consider a *reversible* Turing reduction $f = y \leq_{\mathrm{T}} g = z$, that is, we assume $g = z$ is also reducible to $f = y$. As a result, the **least** number of queries needed by solving $g(x) = z$ is equal to $time(g = z) = \dfrac{\mathrm{I}(g = z)}{\mathrm{I}(g = z; f = y)}$. By the symmetry of the pointwise mutual information, we obtain the following relationship

$$\frac{\mathrm{I}(f = y)}{time(f = y)} = \mathrm{I}(f = y; g = z) = \mathrm{I}(g = z; f = y) = \frac{\mathrm{I}(g = z)}{time(g = z)}.$$

Specially, if $f$ is *Turing equivalent* to $g$, then the average number of queries must be related by

$$\frac{\mathrm{I}(f = y)}{time(f = y)} = \frac{\mathrm{I}(g = z)}{time(g = z)} = \mathrm{I}(f; g).$$

According to the data processing inequality (see section 3.4), $f \equiv_{\mathrm{T}} g$ means that the average mutual information $\mathrm{I}(f; g)$ is **maximal** among all possible Turing reductions from $f(x)$ to other functions, because $f(x)$ and $g(x)$ contain all useful information about each other.

In the ideal case, if $\mathrm{I}(f; g) = \mathrm{H}(f)$, then we obtain a perfect analogy of an ideal heat engine

$$\frac{\mathrm{I}(f = y)}{time(f = y)} = \frac{\mathrm{I}(g = z)}{time(g = z)} = \mathrm{H}(f).$$

In this ideal case, $time(f = y) = \dfrac{\mathrm{I}(f = y)}{\mathrm{H}(f)}$ is **minimal** among all possible Turing reductions from $f(x) = y$ to other functions. *No algorithm can solve $f(x) = y$ with query complexity strictly less than* $\dfrac{\mathrm{I}(f = y)}{\mathrm{H}(f)}$. Once again, this limitation is the property of the nature, not the property of a particular problem.

In conclusion, working always with reversible Turing reductions, the query complexity of solving $f(x) = y$ *does not depend on the design of the Turing machine*. This is a perfect analogy of Carnot's brilliant conclusion: that the efficiency of a reversible heat engine *does not depend on the design of the heat engine* (see Feynman *et al.* [2013]).

If $\dfrac{\mathrm{I}(f = y)}{\mathrm{H}(f)}$ is *exponential* in the size of input, then the search problem $f(x) = y$ can not be reduced to any other function within polynomial time. According to the Church–Turing Thesis, in this case $f(x) = y$ is less likely to have polynomial-time algorithm on any ordinary Turing machine. In fact, according to Markov's inequality, if $\dfrac{I(f = y_k)}{H(f)}$ is exponential in the size of input, then the probability of $f(x) = y$ having polynomial-time algorithm is a negative exponential in the size of input. This result enables us to reduce the problem of proving $P \neq NP$ to the computation of the entropy of the subset sum function (section 5.1). As a result, $P \neq NP$ is provable, answering a long-standing open problem (see S. Aaronson [2003]).

As it turns out, our information-based complexity estimation is a natural setting in which to study the power of randomized or probabilistic algorithms. In fact, our information-based estimation of the lower bound on repetitions of Polynomial-time Probabilistic algorithm is superior to that based on the Chernoff bound (see Papadimitriou [1994]).

In theory, $\dfrac{I(f = y)}{I(f;g)}$ gives rise to an **upper bound** on the query complexity of solving $f(x) = y$. In the idea case, if $\dfrac{I(f = y)}{I(f;g)}$ is polynomial in the size of input and the function $g(x)$ can be computed in polynomial time, then the problem $f(x) = y$ also has polynomial-time algorithm. Applying to decision problems, our result provides strong evidence that $P = RP = BPP$.

Finally, using brute-force search as a benchmark for efficiency, our results support that $P \neq PP$. This result is obtained by estimating the lower bound on the average mutual information for any reduction-based algorithm that is more efficient than brute-force search. However, the existence of such a lower bound contradicts with the definition of the class $PP$.

The main contributions of this paper can be summarized as follows.

— The concept of Shannon entropy of random variables was generalized to measurable functions in general, and to simple functions with finite values in particular.

— It is shown that the information measure of functions is related to the time complexity of solving search problems concerning functions in question.

— By estimating the **least** number of queries, we are able to open the "black box" of oracle Turing machines.

— Within our framework the problem of proving $P \neq NP$ can be reduced to the computation of the entropy of certain simple functions, such as the subset sum function.

— Our information-based complexity estimation provides strong evidence that $P = RP = BPP$.

— Using brute-force search as a benchmark for efficiency, our results support that $P \neq PP$.

## ACKNOWLEDGMENTS

The authors would like to thank Professor Gaoping Li and Doctor Yihong Chen for helpful discussions.